\documentclass[acmtog,nonacm]{acmart}
\acmSubmissionID{}

\usepackage{booktabs} 
\usepackage[leftcaption]{sidecap}

\usepackage[ruled]{algorithm2e} 

\SetAlFnt{\small}
\SetAlCapFnt{\small}
\SetAlCapNameFnt{\small}
\SetAlCapHSkip{0pt}

\acmJournal{TOG}

\usepackage{color}
\usepackage{graphicx}
\usepackage{caption}
\usepackage{wrapfig}
\usepackage{caption}
\usepackage{wrapfig}
\usepackage{wrapfig}
\usepackage{hyperref}
\usepackage{cleveref}
\usepackage{stfloats} 
\usepackage{rotating}
\usepackage{placeins} 
\usepackage{overpic}

\definecolor{title_purple}{rgb}{0.65,0.1,0.65}

\definecolor{green}{rgb}{0, 0.5, 0}
\definecolor{orange}{rgb}{0.8, 0.6, 0.2}
\definecolor{red}{rgb}{1.0, 0.0, 0.0}
\definecolor{teal}{rgb}{0.0, 0.4, 0.4}
\definecolor{purple}{rgb}{0.65,0.0,0.65}
\definecolor{saffron}{rgb}{0.95,0.75,0.2}
\definecolor{turquoise}{rgb}{0.0,0.5,0.5}
\definecolor{black}{rgb}{0.0, 0.0, 0.0}
\definecolor{gray}{rgb}{0.5, 0.5, 0.5}
\definecolor{darkpink}{rgb}{0.561, 0.282, 0.427}
\definecolor{darkturquoise}{rgb}{0.4, 0.52, 0.83}

\newcommand{\sm}[1]{{\color{darkturquoise}#1}}

\makeatletter
\DeclareRobustCommand\onedot{\futurelet\@let@token\@onedot}

\makeatother

\makeatletter
\def\blfootnote{\xdef\@thefnmark{}\@footnotetext}
\makeatother

\begin{document}

\title{Untwisting RoPE: Frequency Control for Shared Attention in DiTs}

\author{Aryan Mikaeili}
\affiliation{%
 \institution{Simon Fraser University}
 \city{Burnaby}
 \country{Canada}}
\email{aryan_mikaeili@sfu.ca}
\author{Or Patashnik}
\affiliation{%
 \institution{Tel Aviv University}
 \city{Tel Aviv}
 \country{Israel}
}
\author{Andrea Tagliasacchi}
\affiliation{%
 \institution{Simon Fraser University,}
 \institution{University of Toronto}
 \country{Canada}
}
\author{Daniel Cohen-Or}
\affiliation{%
 \institution{Tel Aviv University}
 \city{Tel Aviv}
 \country{Israel}
}
\author{Ali Mahdavi-Amiri}
\affiliation{%
 \institution{Simon Fraser University}
 \city{Burnaby}
 \country{Canada}}



\begin{abstract}
Positional encodings are essential to transformer-based generative models, yet their behavior in multimodal and attention-sharing settings is not fully understood. In this work, we present a principled analysis of Rotary Positional Embeddings (RoPE), showing that RoPE naturally decomposes into frequency components with distinct positional sensitivities. We demonstrate that this frequency structure explains why shared-attention mechanisms, where a target image is generated while attending to tokens from a reference image, can lead to reference copying, in which the model reproduces content from the reference instead of extracting only its stylistic cues. Our analysis reveals that the high-frequency components of RoPE dominate the attention computation, forcing queries to attend mainly to spatially aligned reference tokens and thereby inducing this unintended copying behavior.
Building on these insights, we introduce a method for selectively modulating RoPE’s frequency bands so that attention reflects semantic similarity rather than strict positional alignment. Applied to modern transformer-based diffusion architectures, where all tokens share attention, this modulation restores stable and meaningful shared attention. As a result, it enables effective control over the degree of style transfer versus content copying, yielding a proper style-aligned generation process in which stylistic attributes are transferred without duplicating reference content.
\end{abstract}
\begin{teaserfigure}
\centering
  \includegraphics[width=0.9\textwidth]{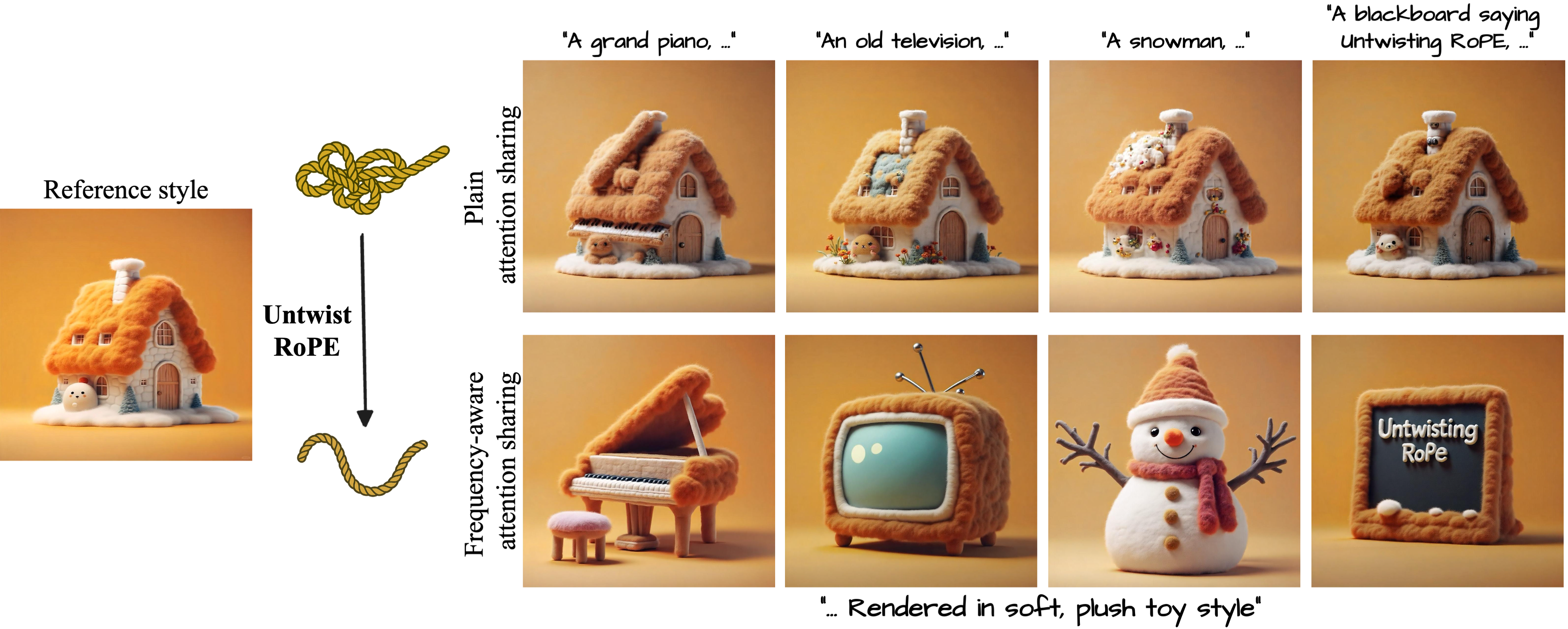}
  \caption{Shared-attention in RoPE-based diffusion transformers often collapses into reference copying: high-frequency RoPE components strongly bias attention toward spatially aligned reference tokens, causing the model to reproduce reference content rather than extract its style \textbf{(top row)}. We analyze this frequency structure and introduce a frequency-aware modulation of RoPE that restores meaningful, semantically guided shared attention. Our method enables controllable style-aligned generation in which stylistic attributes are transferred without duplicating reference content \textbf{(bottom row)}.}
  \Description{}
  \label{fig:teaser}
\end{teaserfigure}
%
%

\maketitle

\section{Introduction}
\label{ref:intro}


Transformers underpin today’s most powerful generative models~\cite{Peebles2022DiT, flux2024}.
In particular, diffusion transformers, such as Multi-modal Diffusion Transformers~(MMDiTs), adapt the transformer architecture to the diffusion framework, enabling joint modeling of textual and visual representations throughout the denoising process.
For these models, positional encodings are essential, as they supply the spatial structure that the attention mechanism itself lacks. 
Without positional information, the model treats tokens as an unordered set, preventing coherent reasoning about spatial layout and making high-quality image generation impossible. Among the various designs, Rotary Positional Embeddings (RoPE)~\cite{su2021roformer} have proven effective in practice and are widely adopted in diffusion transformers, injecting relative position information directly into attention. RoPE therefore serves as a key mechanism that restores the locality once provided by convolutional networks.

Beyond standard text-conditioned generation, shared-attention mechanisms have proven highly effective in UNet-based diffusion models, where tokens from a target image attend directly to those of a reference. This design provides a flexible interface for image manipulation tasks such as appearance transfer~\cite{alaluf2023crossimage}, style transfer~\cite{hertz2023StyleAligned, Deng_2024_CVPR, deng2024z}, and reference-based editing~\cite{cao_2023_masactrl, tokenflow2023}, by implicitly computing semantic correspondences between images. As diffusion transformers increasingly replace convolutional UNets, understanding how these shared-attention mechanisms translate to transformer-based architectures becomes of critical importance.

\begin{figure}[t!]
    \centering
    \begin{overpic}[width=0.85\linewidth]
    {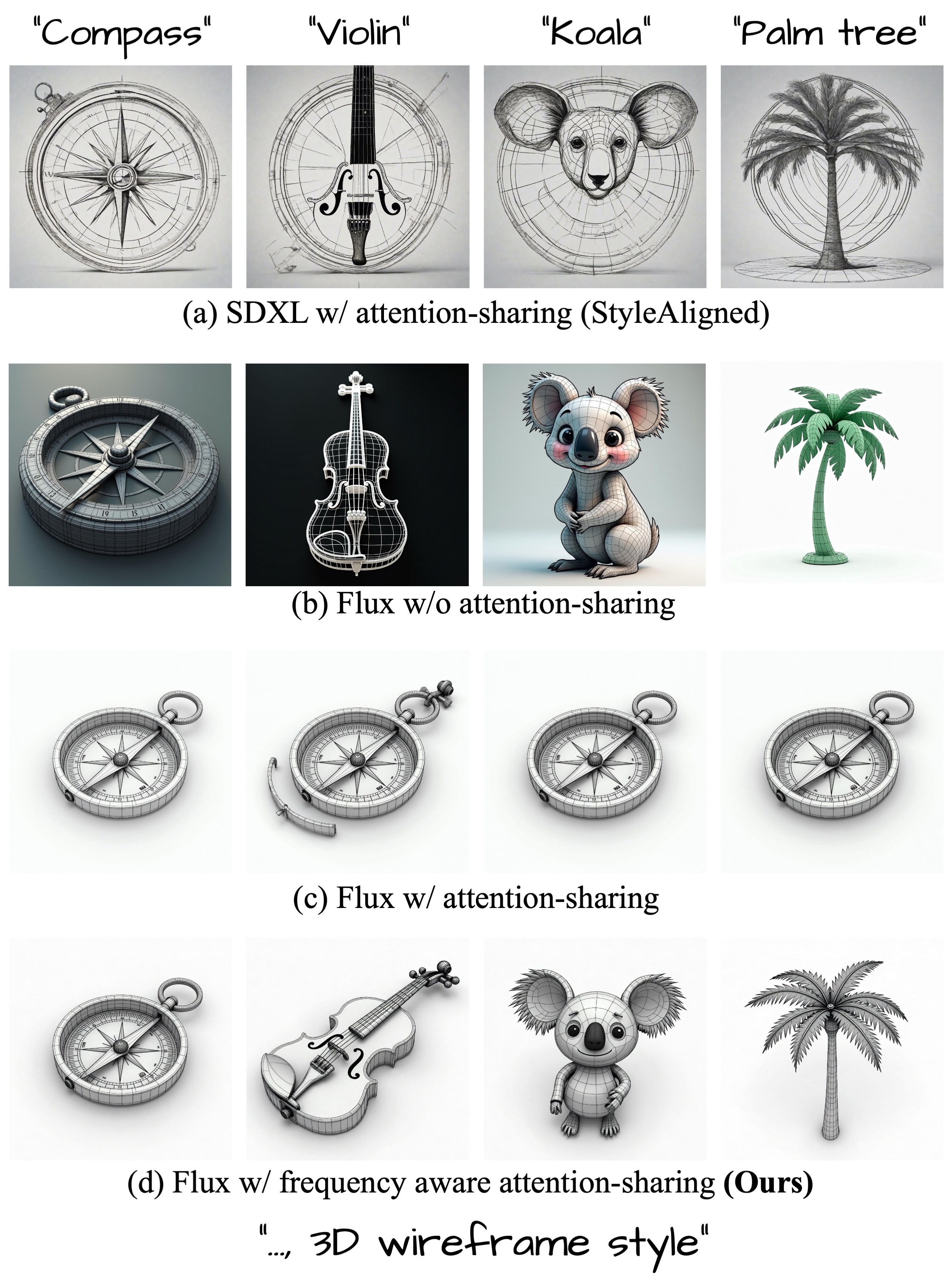}
    \end{overpic}
    \caption{\textbf{Style-aligned image generation} aims to produce images with different contents but a consistent style.
    \textbf{(a)} StyleAligned~\cite{hertz2023StyleAligned} applies shared attention in SDXL, achieving style alignment but introducing artifacts.
    \textbf{(b)} Flux without attention sharing fails to produce style-aligned image sets.
    \textbf{(c)} Plain attention sharing in Flux leads to near-identical outputs due to positional bias.
    \textbf{(d)} Our frequency-aware attention sharing produces style-aligned images while preserving content fidelity to the text prompts.
}
\label{fig:summary}
\end{figure}

Specifically, unlike UNet–based diffusion models, whose attention layers do not rely on positional encodings, diffusion transformers depend critically on positional information, with RoPE playing a central role.
In shared-attention settings, however, the strong spatial bias introduced by RoPE can dominate the attention computation, disrupting the semantic correspondences that shared attention relies on. As a result, instead of enabling stylistic or appearance transfer, the model often exhibits \textit{reference copying}, reproducing content from the reference image rather than transferring features from it~(see \Cref{fig:teaser}).
This effect has been recently observed~\cite{Avrahami_2025_CVPR, zhang2025alignedgen, wei2025freeflux, chen2025devil}, but it has not yet been clearly understood.
Style-aligned image generation aims to produce multiple images with diverse semantic content while maintaining a consistent visual style. As illustrated in Figure \ref{fig:summary}, when this objective is pursued in diffusion transformers, either disabling attention sharing fails to align styles, or enabling it naïvely leads to collapse into near-identical images due to content leakage from the reference.

To understand this behavior, we perform a principled analysis of RoPE and study its impact on the shared attention mechanism. Specifically, we show that different frequency components of RoPE exhibit different degrees of positional sensitivity. 
In particular, we show that the high-frequency components exert a disproportionately strong influence on the attention computation, steering queries in the target image toward spatially aligned reference tokens. This positional dominance provides a clear explanation for why shared-attention mechanisms tend to collapse into content copying (see \Cref{fig:teaser}-top row).

Building on this insight, we introduce an effective method for selectively modulating the frequency bands of RoPE. By attenuating the high-frequency components or amplifying the low-frequency ones, we rebalance the positional bias and allow attention to be guided by semantic similarity rather than strict spatial alignment. When applied to MMDiT-based architectures, this modulation restores stable and meaningful shared attention (\Cref{fig:teaser}-bottom).

Figure~\ref{fig:summary} summarizes the problem setting and the resulting contribution of this work. While shared attention enables style-aligned generation in UNet-based diffusion models (\Cref{fig:summary}-a), directly applying it to diffusion transformers leads to systematic failures: without attention sharing, styles do not align (\Cref{fig:summary}-b), and naïve sharing collapses the generation into near-identical images due to content leakage (\Cref{fig:summary}-c). We show that this behavior stems from RoPE’s positional bias and that selectively modulating its frequency components restores meaningful shared attention, enabling style-aligned generation that preserves semantic diversity (\Cref{fig:summary}-d).

This understanding enables practical and controllable style-aligned generation: the model can follow stylistic cues from a reference image without mistakenly reproducing its content.
Beyond style-aligned generation, our analysis reveals that the frequency structure of RoPE provides a direct and controllable handle for shaping how attention trades off between positional locality and semantic association in diffusion transformers.

\section{Related Works}

\paragraph{\textbf{Attention-based image manipulation}}
Attention sharing allows target tokens to attend to reference tokens during synthesis and has been explored across image generation and editing. Plug-and-Play~\cite{Tumanyan_2023_CVPR} injects reference attention for editing, MasaCtrl~\cite{cao_2023_masactrl} replaces target keys and values to enable non-rigid edits, and Alaluf et al.~\shortcite{alaluf2023crossimage} show that attention sharing generalizes across images for zero-shot appearance transfer.

This mechanism has also been applied to localized image editing~\cite{patashnik2024qnerf, almohammadi2025coracorrespondenceawareimageediting, xu2023infedit, koo2024flexiedit, Avrahami_2025_CVPR}, layout-guided generation~\cite{eldesokey2025buildascene, generativephotomontage, mikaeili2025griffingenerativereferencelayout}, identity-consistent synthesis~\cite{tewel2024consistory}, and video editing~\cite{tokenflow2023, qi2023fatezero}.

StyleAligned~\cite{hertz2023StyleAligned} is a representative shared-attention method for generating style-consistent image sets. However, it is built on UNet architectures, whose attention formulation and convolutional inductive biases differ fundamentally from those of DiTs. This mismatch complicates the transfer of shared-attention techniques to transformer-based diffusion models, where positional structure is encoded exclusively via RoPE.


\paragraph{\textbf{Diffusion transformers and RoPE}}

Diffusion Transformers (DiTs)~\cite{Peebles2022DiT} have become the dominant architecture for text-to-image generation~\cite{flux2024, Esser2024SD3}, largely replacing UNet-based designs~\cite{Rombach2022LDM, Podell2024SDXL}. By jointly processing image patches and text tokens within a unified transformer backbone, DiTs enable scalable models with strong image--text alignment, high-resolution synthesis, and flexible conditioning via additional context tokens~\cite{labs2025flux1kontextflowmatching, wu2025qwenimagetechnicalreport}.

Because transformers lack inherent spatial awareness, positional encodings are required to impose structure. Early vision transformers used sinusoidal encodings~\cite{vaswani2017attention, dosovitskiy2020vit}, while modern DiT-based models adopt rotary positional encodings (RoPE)~\cite{su2021roformer}, which encode relative positions and remain robust to changes in spatial and temporal resolution~\cite{flux2024, labs2025flux1kontextflowmatching, wan2025}.
Few works exploit properties of RoPE for specialized tasks such as high resolution image generation~\cite{Issachar2025DyPE} by \textit{rescaling}~\cite{Chen2023PositionalInterpolation, Reddit2023NTKAwareRoPE, Peng2024YaRN}, or view synthesis~\cite{bai2025positional} by \textit{warping} positional encodings. 

Other works note that attention sharing in RoPE-based DiTs can lead to unintended content copying. Methods such as~\cite{wei2025freeflux, wang2024taming, Avrahami_2025_CVPR} mitigate this by restricting attention sharing to selected DiT blocks, providing only coarse control over reference influence. 
A concurrent work, AlignedGen~\cite{zhang2025alignedgen}, proposes shifting the positional indices of reference tokens to reduce positional collisions.
While effective, shifting offers limited controllability and can introduce artifacts such as \emph{ghost-cyclic inpainting} (see Figure~\ref{fig:visual_comparison}).
This stems from a lack of systematic analysis of RoPE and an explanation of why shared attention in DiTs collapses into content copying.
In this paper, we address this gap and derive a more controllable and general solution.



\textit{\textbf{Style transfer}}
has a long history in computer graphics~\cite{hertzmann2001analogies, efros2023quilting,Gatys2016StyleTransfer, huang2017adain,Johnson2016PerceptualLosses,pix2pix2017, CycleGAN2017, huang2018munit}.
Recent style-transfer works extend text-to-image diffusion models for controllable style transfer. Personalization-based methods such as textual inversion~\cite{gal2022textual}, DreamBooth~\cite{ruiz2022dreambooth}, StyleDrop~\cite{Sohn2023StyleDrop}, and B-LoRA fine-tune model parameters from a small set of examples to capture a style or subject~\cite{frenkel2024implicit}.

Encoder-based approaches condition diffusion models on reference images via learned style encoders, including IP-Adapter~\cite{ye2023ip-adapter}, Instant-Style~\cite{wang2024instantstyle}, Instant-Style Plus~\cite{wang2024instantstyleplus}, StyleCrafter~\cite{liu2023stylecrafter}, and StyleMaster~\cite{ye2025stylemaster}. However, these methods are limited to styles seen during training, and often require retraining for each backbone.
Training-free approaches based on shared or cross-image attention offer an alternative. Cross-image attention~\cite{alaluf2023crossimage} and StyleAligned~\cite{hertz2023StyleAligned} demonstrate effective attention-driven style transfer in UNet-based diffusion models, highlighting attention as a powerful mechanism for style control—an idea our work further explores in the context of DiTs.

\section{Preliminaries}  
\begin{figure}[ht]
    \centering
    \begin{overpic}[width=\linewidth]
    {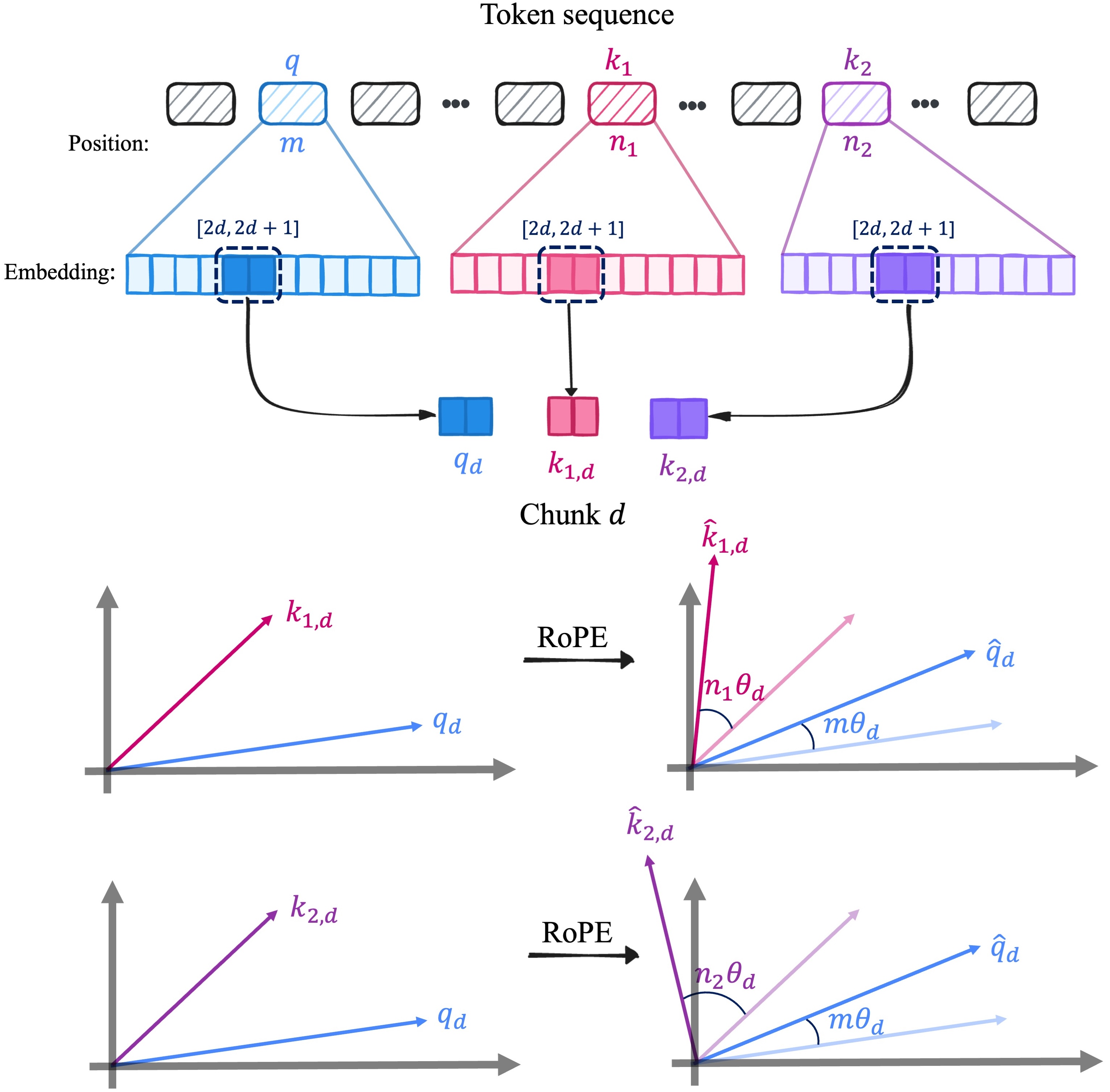}
    \end{overpic}
    \put(-108, 105){\scriptsize{(a)}}
    \put(-108, -5){\scriptsize{(b)}}
    \caption{\textbf{RoPE overview.}
(a) For a token sequence, we illustrate three tokens, $q$, $k_1$, and $k_2$, located at positions $m$, $n_1$, and $n_2$.
For each token, we highlight three corresponding \emph{chunks}: 2-dimensional slices of the embedding on which RoPE applies independent rotations.
(b) Assuming the corresponding chunks of $k_1$ and $k_2$ are identical, RoPE rotates them by different angles because their positions $n_1$ and $n_2$ differ. Each chunk is rotated by frequency $\theta_d$, producing position-dependent inner products between $q$ and each key.
}
\vspace{-10pt}
    \label{fig:rope_overview}
\end{figure}

Recent text-to-image models such as Flux~\cite{flux2024} adopt a multimodal diffusion transformer (MMDiT) architecture~\cite{Peebles2022DiT}, where visual latents and text embeddings are concatenated and processed jointly via multimodal self-attention.
This contrasts with UNet designs, where convolution performs local feature extraction, cross-attention injects text conditioning, and image self-attention provides global image context.
By unifying these interactions in transformer blocks, MMDiTs enable fine-grained, bidirectional exchange between image and text tokens at all layers.

\newcommand{\img}{\text{img}}
\newcommand{\txt}{\text{txt}}
At diffusion timestep $t$, noisy image tokens $H^{\img}_{t}\in\mathbb{R}^{N\times D}$ and text tokens $H^{\txt}\in\mathbb{R}^{M\times D}$ are projected into queries, keys, and values:
\begin{equation}
\begin{aligned}
    Q^{\img} &= W_Q^{\img} H^{\img}_{t}, \: 
    &K^{\img} &= W_K^{\img} H^{\img}_{t}, \: 
    &V^{\img} &= W_V^{\img} H^{\img}_{t},
    \\
    Q^{\txt} &= W_Q^{\txt} H^{\txt}, \:
    &K^{\txt} &= W_K^{\txt} H^{\txt}, \:
    &V^{\txt} &= W_V^{\txt} H^{\txt}.
\end{aligned}
\end{equation}
After projection\footnote{Flux uses two attention block types: \emph{single-stream} blocks (shared projections across modalities where $W_{Q,K,V}^{\img} = W_{Q,K,V}^{\txt}$) and \emph{dual-stream} blocks (separate projections). Our attention-sharing is applied only to single-stream blocks.}, image and text tokens are concatenated and the attention output is
\begin{align}
    O = A \cdot [V^{\img}\oplus V^{\txt}],
\end{align}
where $\oplus$ denotes concatenation and the attention matrix $A$ is
\begin{align}
     A = \text{Softmax}\!\left(\frac{\text{RoPE}([Q^{\img}\oplus Q^{\txt}])^\top\cdot\text{RoPE}([K^{\img}\oplus K^{\txt}])}{\sqrt{D}}\right),
\end{align}
with RoPE denoting rotary positional encoding. In the following, we analyze how RoPE shapes attention.


\subsection{Rotary Positional Encoding (RoPE)}
\label{subsec:rope}
The attention operation is inherently permutation-equivariant. To model the strong spatial correlations in images, positional information must therefore be explicitly injected. Modern text-to-image models such as Flux~\cite{flux2024} achieve this by applying Rotary Positional Embeddings (RoPE)~\cite{su2021roformer} to the queries and keys in each transformer block, enabling the model to encode relative spatial relationships among tokens.

\noindent
\paragraph{\textbf{One-dimensional sequence}}
Given a one-dimensional token sequence, let $q \in \mathbb{R}^D$ denote the query at position $m$,   and $k \in \mathbb{R}^D$ the key at position $n$.
Let us understand how the relative displacement $(n-m)$ between query and key affects attention.
RoPE partitions each $D$-dimensional vector into $D/2$ \textit{chunks}, indexed by $d$, each containing a pair of consecutive vector entries, and applies two-dimensional rotations element-wise (see Figure~\ref{fig:rope_overview}):
\[
    \hat{q}_d = R_{m\theta_d}\, q_d, \qquad 
    \hat{k}_d = R_{n\theta_d}\, k_d, 
\]
with $\theta_{\text{base}} = 1/10000$ as default:
\[
\theta_d = \theta_{\text{base}}^{\,2d/D}, 
\qquad 
d \in \left\{0,\dots,\tfrac{D}{2}-1\right\}.
\]


The attention between $q$ and $k$ is then proportional to the inner
product of their rotated embeddings:
\begin{equation}
\begin{aligned}
    A_{q \rightarrow k} = A[q, k]
    &\propto \langle \hat{q}, \hat{k} \rangle=\\
 \sum_{d=0}^{D/2 - 1} q_d^\top R_{m\theta_d}^\top R^{\vphantom{\top}}_{n\theta_d} k_d 
    &= \sum_{d=0}^{D/2 - 1} q_d^\top R_{(n-m)\theta_d} k_d
    = \sum_{d=0}^{D/2 - 1} \langle q_d, R_{(n-m)\theta_d}\, k_d \rangle. \nonumber
\end{aligned}
\end{equation}
revealing that RoPE encodes positions through a relative rotation
proportional to the \textit{displacement} $(n - m)$.

An overview is shown in \Cref{fig:rope_overview}, illustrating that a query at position $m$ attends differently to keys at positions $n_1$ and $n_2$ due to distinct rotations of their two-dimensional chunks.


\begin{figure*}[t]
    \centering
    \begin{overpic}[width=0.9\linewidth]
    {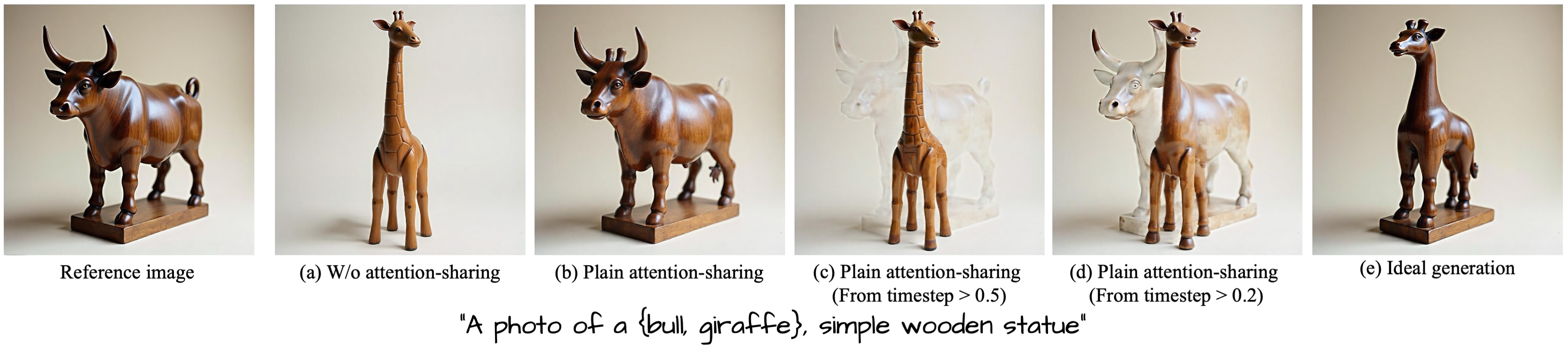}
    \end{overpic}
       \caption{\textbf{Reference copying.} Given a reference image of a bull, we generate a style-aligned image of a giraffe.
        \textbf{(a)} Without attention sharing, the output is not style-aligned.
        \textbf{(b)} Applying attention sharing at all denoising timesteps leads to near-exact replication of the reference.
        \textbf{(c,d)} Restricting attention sharing to later timesteps does not prevent copying.
        \textbf{(e)} Our method produces a style-aligned result while preserving the target prompt.
   }
    \label{fig:copying}
\end{figure*}

\paragraph{\textbf{Multi-dimensional sequence}}
For multi-dimensional sequences such as images, each $D$-dimensional token is split into two sub-vectors that encode positional dependencies along the $x$- and $y$-axes, allowing RoPE to model each spatial dimension independently. 
In models such as Flux and in video generative models, an additional portion of the embedding encodes temporal positional dependencies; in Flux, this temporal component is kept unrotated.
In the calculation of the attention of a query at position~$(x_q, y_q)$ to a key at position~$(x_k, y_k)$ for the 2D chunk of each token corresponding to the $x$-axis~($d_x$), the 
attention inner product becomes
\begin{align}
    \langle \hat{q}_{d_x}, \hat{k}_{d_x} \rangle 
    &= \langle q_{d_x},\, R_{(x_k - x_q)\theta_{d_x}}\, k_{d_x} \rangle,
    d_x \in \left\{0, \dots, \tfrac{D_x}{2}-1 \right\},
    \label{eq:inner_x}
\end{align}
and along the $y$-axis, the inner product is calculated similarly.
Finally, it is also worth noting that text tokens are assigned position zero.

\section{Attention-sharing} 
\label{sec:study}

Attention sharing has been widely studied in UNet-based text-to-image models for
image manipulation tasks~\cite{hertz2023StyleAligned, tewel2024consistory,
mou2023dragondiffusion}. In this work, we focus on style-aligned image generation, introduced in StyleAligned~\cite{hertz2023StyleAligned}, and briefly review the method.

StyleAligned modifies UNet self-attention so that target image tokens attend to reference tokens during denoising.
The method applies AdaIN~\cite{huang2017adain} to the target queries and keys, then concatenates the reference keys and values with those of the target before computing self-attention, enabling extraction of style cues via shared attention.
%
\begin{equation}
\begin{aligned}
    Q &= \tilde{Q}_\text{tar}, \quad &\tilde{Q}_\text{tar} &= \text{AdaIN}(Q_\text{tar}, Q_\text{ref}), \\
    K &= \tilde{K}_\text{tar} \oplus s \cdot K_\text{ref}, \quad & \tilde{K}_\text{tar} &= \text{AdaIN}(K_\text{tar}, K_\text{ref}), 
    \\
    V &= V_\text{tar} \oplus V_\text{ref},
\end{aligned}
\end{equation}
where $\{Q,K,V\}_{ref,tar}$ denote reference and target tokens, and $s$ is a scalar controlling
the strength of style transfer.
By adding RoPE this formulation can be extended to MMDiTs by concatenating reference image tokens with the target image keys and values:
\begin{equation}
\begin{aligned}
    Q &= \text{RoPE}([\tilde{Q}^{\text{img}}_\text{tar} \oplus Q^{\text{txt}}_\text{tar}]), \label{eq:mmdit_sharing} \\
    K &= \text{RoPE}([\tilde{K}^{\text{img}}_\text{tar} \oplus K^{\text{txt}}_\text{tar}]) \oplus s \cdot \text{RoPE}(K^{\img}_\text{ref}), \\
    V &= [V^{\img}_\text{tar} \oplus V^{\txt}_\text{tar}] \oplus V^{\img}_\text{ref}.
\end{aligned}
\end{equation}
However, directly applying this formulation leads to \textit{reference copying} rather than style transfer, as discussed next.

\textbf{\textit{Reference copying}} is a failure mode in which the generated image reproduces visual content from the reference image rather than its stylistic cues.
As shown in \Cref{fig:copying}-b, instead of producing a style-aligned version of the target prompt (giraffe), the model ``collapses'' and duplicates shapes, colors, and fine details from the reference (bull). Further, although prior work suggests that stylistic and appearance details emerge at later denoising steps and content is formed at earlier steps~\cite{Issachar2025DyPE, bahmani2025ac3d}, restricting attention sharing to later stages does not resolve the issue entirely (\Cref{fig:copying}-c,d).

This behavior was also noticed by~\citet{zhang2025alignedgen} and \citet{wei2025freeflux}, and attributed to RoPE disproportionately amplifying attention to spatially aligned reference tokens.
During denoising, this repeated alignment draws visual information from identically positioned reference tokens, resulting in content copying. In the following, we analyze the underlying cause of this behavior through the lens of RoPE frequency bands.

\begin{SCfigure*}[0.6][t]
    \centering
    \includegraphics[width=1.5\linewidth]{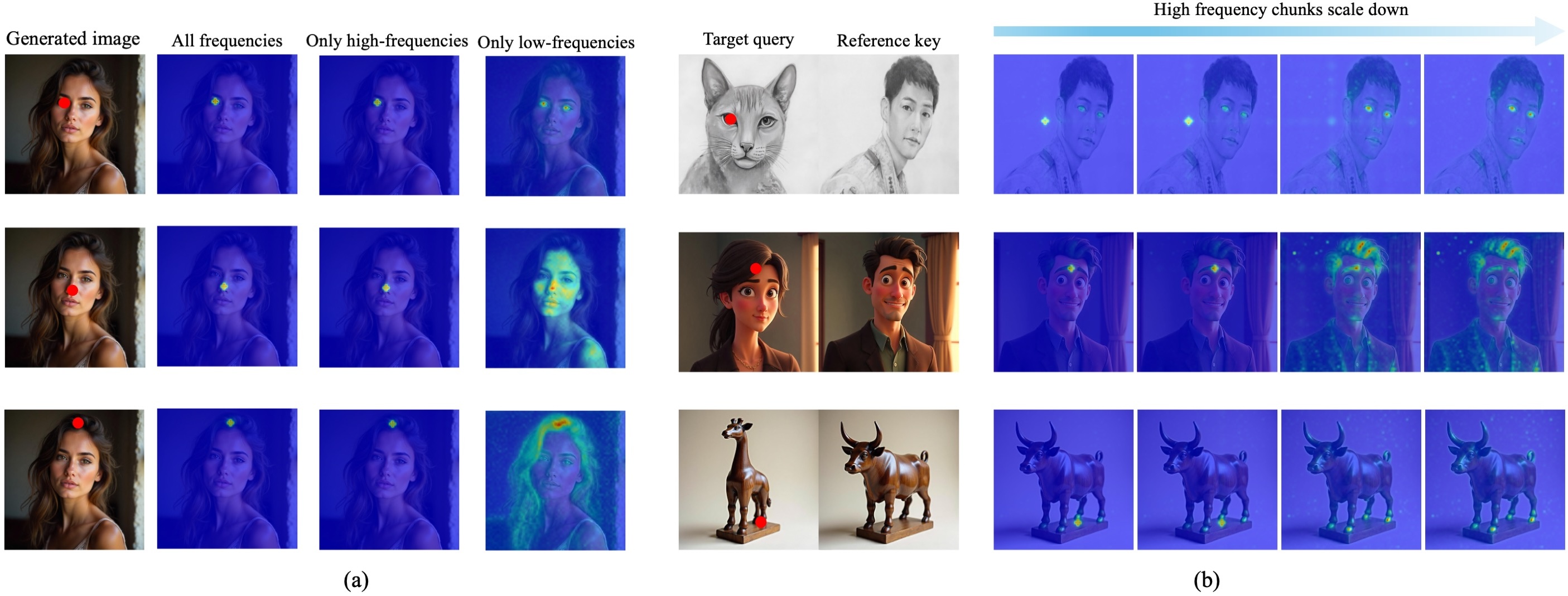}
     \caption{\textbf{(a) Attention visualization for standard image generation.}
        For each query point (red dots), we show the full attention map, attention using only high-frequency RoPE components, and attention using only low-frequency components (left to right). Retaining only low-frequency components yields more semantically aligned attention.
        \textbf{(b) Cross-image attention visualization.}
        Target queries attend to reference image patches; scaling down high-frequency key components similarly makes attention more global.
   }
     \label{fig:attention_vis}
\end{SCfigure*}

\paragraph{\textbf{RoPE frequency bands and attention}}
\label{sec:rope_freq}
The reference copying described above stems from how attention maps are formed in DiTs and how they are influenced by RoPE, which we analyze here. The attention inner product in \Cref{eq:inner_x}
can be rewritten as:
\begin{align}
\langle \hat{q}_d, \hat{k}_d \rangle
=
\langle q_d, R_{\Delta \theta_d} k_d \rangle,
\end{align}
where $\Delta$ denotes the relative positional difference along the spatial axis
corresponding to the given RoPE chunk, i.e.,
$\Delta = x_q - x_k$ for $x$-axis chunks and
$\Delta = y_q - y_k$ for $y$-axis chunks.

Writing each two-dimensional RoPE chunk in polar coordinates, the rotation
$R_{\Delta \theta_d}$ acts as an additive phase shift in angle space.
Consequently, the inner product between the rotated key and the query can be
expressed as the product of their magnitudes and the cosine of the sum of the
original angular difference and the RoPE-induced rotation:
\begin{align}
    \langle \hat{q}_{d}, \hat{k}_{d} \rangle
    &= \lVert \hat{q}_{d} \rVert\, \lVert \hat{k}_{d} \rVert\,
       \cos\!\left(\alpha_{d} + \Delta\,\theta_{d}\right),
\end{align}
where $\alpha_d$ denotes the angle between the original (unrotated)
chunks $q_d$ and $k_d$. For simplicity, we omit the $x$ and $y$ subscripts.

The frequency term $\theta_{d}$ follows a geometric series 
$\{1, \dots, \frac{1}{10000}\}$.
Thus, lower-index dimensions (larger $\theta_{d}$-high frequency dimensions) produce rapid angle changes 
with respect to $\Delta_{k,q}$ and therefore introduce strong positional 
sensitivity, while higher-index dimensions (smaller $\theta_{d}$-low frequency dimensions) vary slowly 
and contribute weak positional bias, focusing on global similarity.

\begin{figure}[h]
    \centering
    \includegraphics[width=\linewidth]{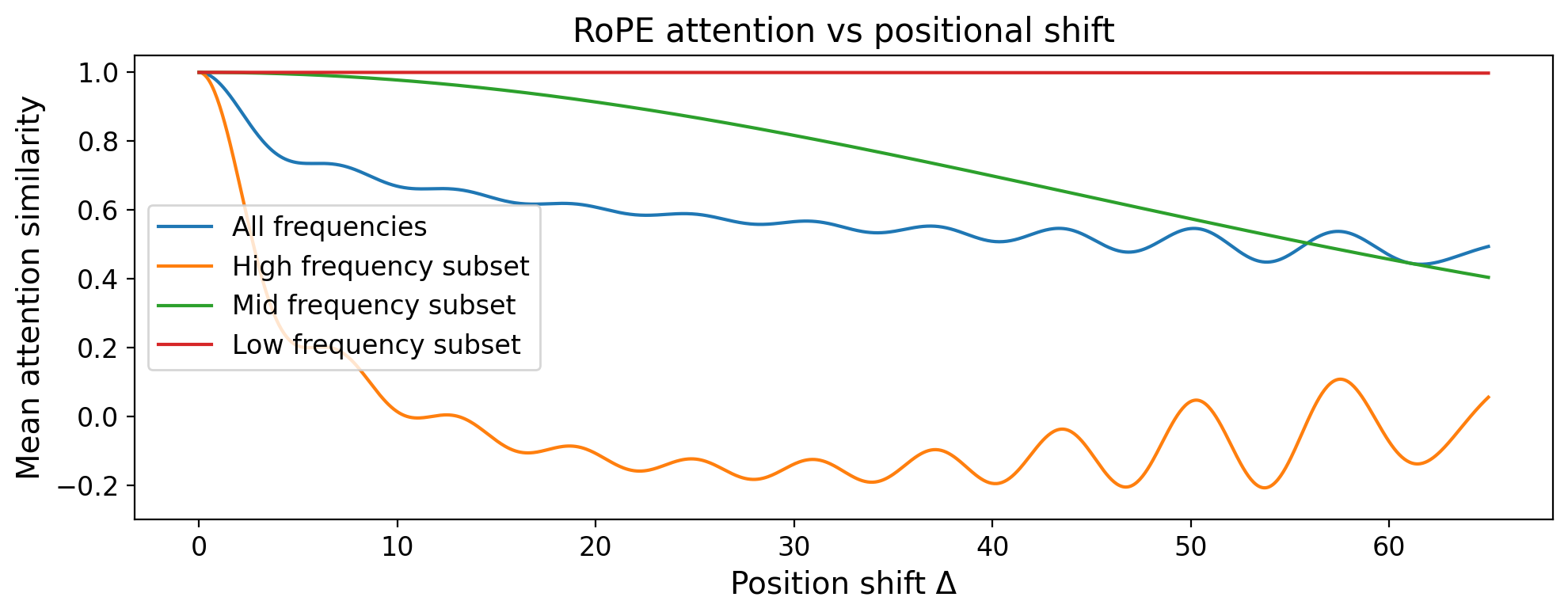}
     \caption{\textbf{Mean attention similarity between two identical vectors as a 
    function of position shift ($\Delta$).} Although RoPE naturally reduces attention 
    similarity as the positional shift increases, the rate of decay varies across 
    frequency bands: high-frequency components exhibit a steep drop, mid-frequency 
    components show moderate sensitivity, and low-frequency components are largely 
    insensitive to positional changes.}
    \label{fig:rope_plot}
    \vspace{-10pt}
\end{figure}

This behavior is further illustrated in \Cref{fig:rope_plot}, where we plot the
mean attention similarity between two identical vectors as a function of their
relative positional difference~$\Delta$. The similarity is defined as
$\cos(\Delta \theta_d)$, corresponding to the normalized inner product of each
two-dimensional RoPE chunk. We report this quantity when all frequency components
are used, and when the vectors are evenly partitioned into high-, mid-, and
low-frequency subsets, with the mean computed separately for each subset.

While RoPE naturally induces decay as $\Delta$ increases, the rate 
of decay varies significantly across frequency bands: the \emph{high-frequency} 
dimensions exhibit a steep drop in similarity even for small $\Delta$, whereas 
the \emph{low-frequency} dimensions remain largely insensitive to $\Delta$. This demonstrates that high-frequency RoPE components enforce strong 
locality, while low-frequency components preserve global spatial coherence.

From this experiment, we expect a similar behavior in image generation with Flux. In \Cref{fig:attention_vis}-a, we separate RoPE features into high- and low-frequency components and average attention across single-stream blocks at timesteps 5, 10, 15, and 20 after zeroing out each component. When high-frequency components are present, attention is strongly biased toward positionally aligned tokens; removing them makes attention more global, with queries attending to semantically similar regions rather than strict spatial alignment.

\begin{figure}
    \centering   
    \includegraphics[width=0.9\linewidth]{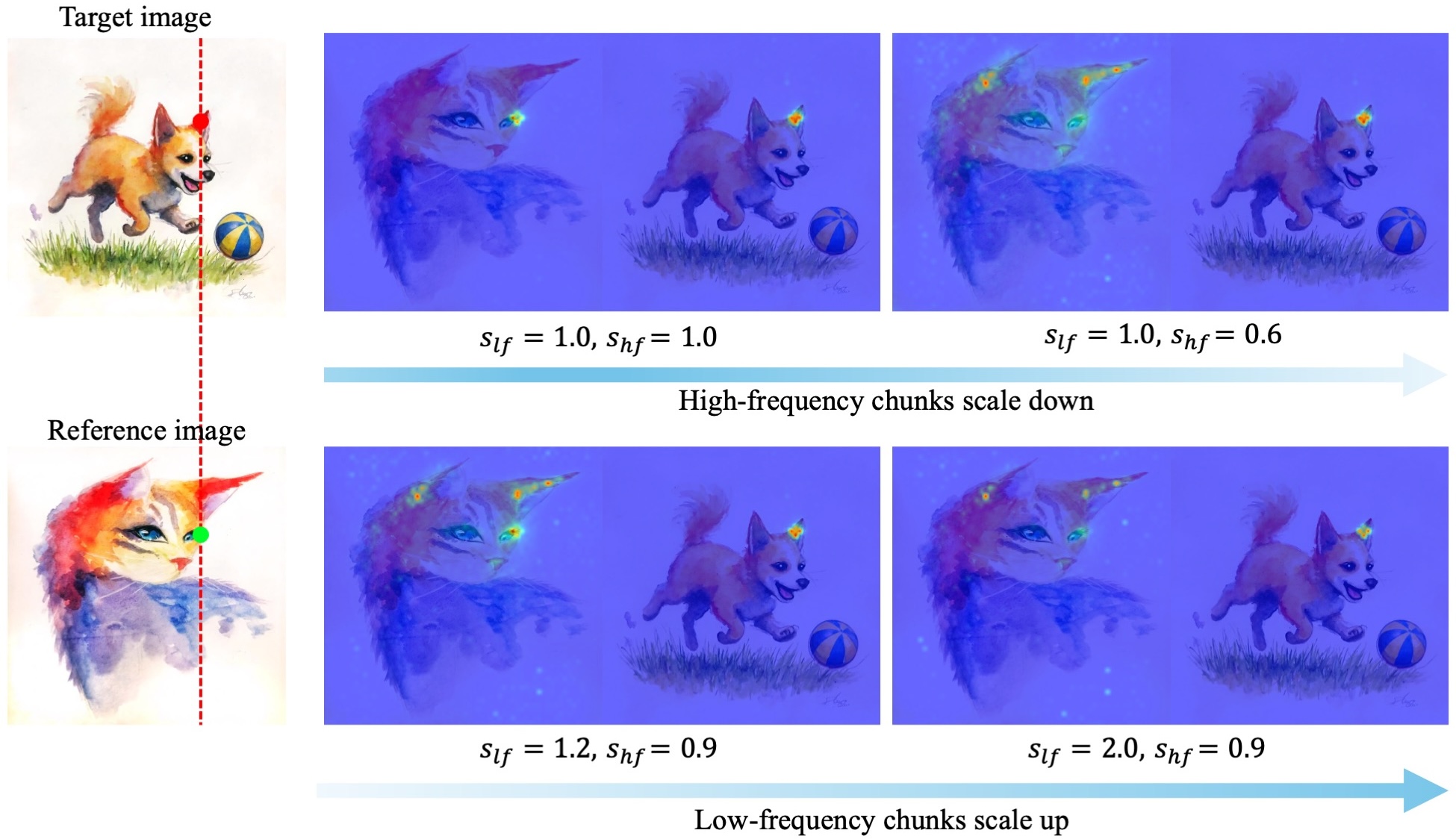}
    \caption{\textbf{Attention visualization for image generation with shared attention.}
    We show attention from a target query (red dot) to both target and reference tokens; the corresponding reference location is marked in green. \textbf{Top:} Scaling down high-frequency components of the reference keys makes attention global and semantically aligned. \textbf{Bottom:} Scaling up low-frequency components has a similar effect.}
    \label{fig:aligned_attention}
\end{figure}

We also analyze cross-image attention, where target queries attend to tokens of another image, to see whether the same behavior is observed. As shown in \Cref{fig:attention_vis}-b, scaling down the high-frequency components of the keys similarly shifts attention toward semantic alignment. This behavior is also reflected in \emph{style-aligned generation}, where reference and target tokens are concatenated (\Cref{fig:aligned_attention}): reducing high-frequency components or amplifying low-frequency ones shifts attention toward semantically meaningful regions (e.g., the cat's ear) rather than positionally aligned tokens.






\sm{
}

\section{Frequency-aware modulation}

\begin{figure}[t]
    \centering
\includegraphics[width=\linewidth]{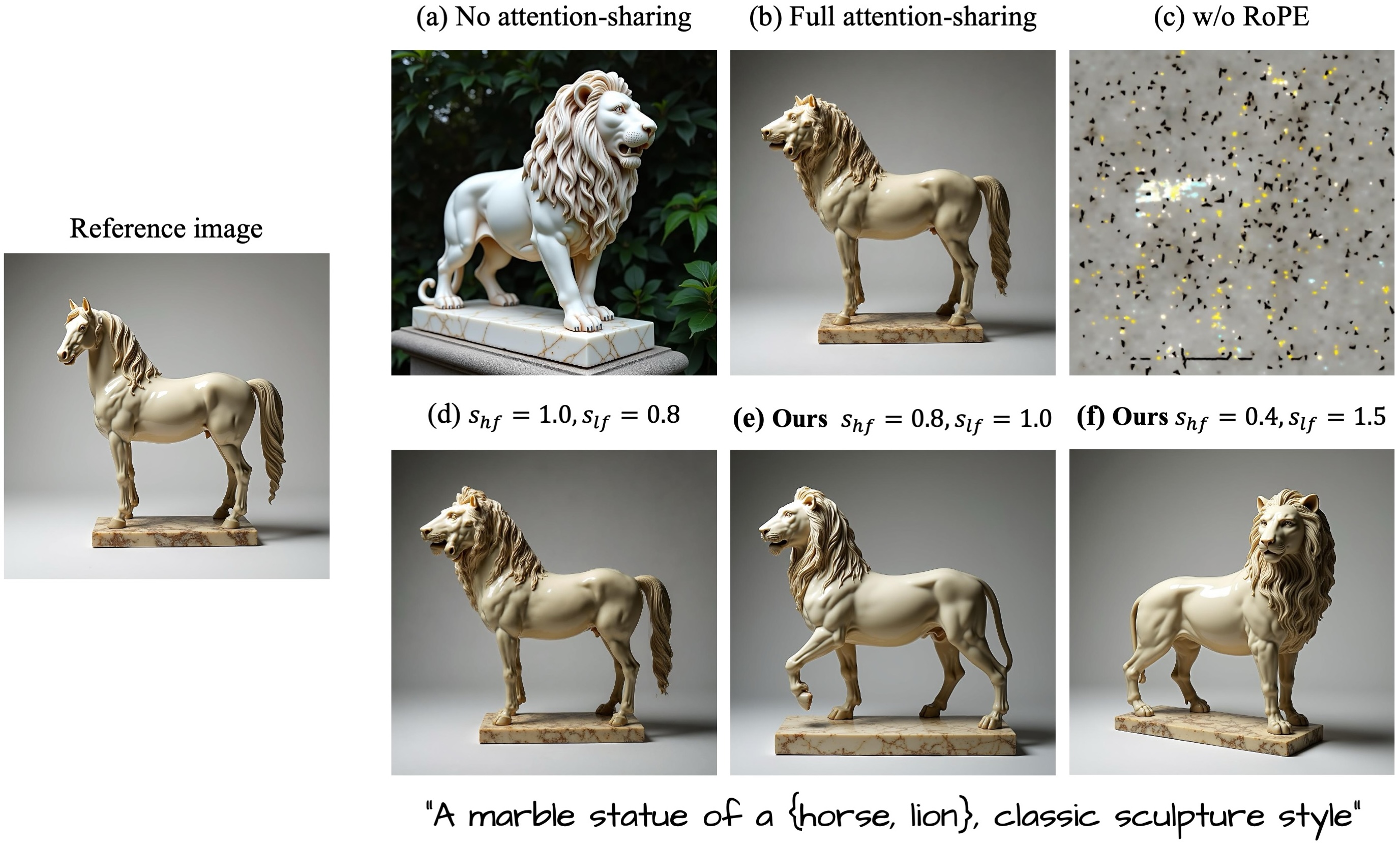}
    \caption{\textbf{Attention sharing in DiTs.}
    \textbf{(a)} Without attention sharing, generated images exhibit inconsistent styles.
    \textbf{(b)} Plain attention sharing causes content copying from the reference.
    \textbf{(c)} Bypassing RoPE during reference attention leads to degenerate generations, showing that positional encoding is essential.
    \textbf{(d)} Scaling down only low-frequency components does not prevent copying, indicating that copying is driven by high-frequency channels.
    \textbf{(e,f)} Suppressing high-frequency components while amplifying low-frequency ones resolves copying and enables balanced style transfer while preserving target semantics.
}
    \label{fig:method_abl}
    \vspace{-10pt}
\end{figure}

In \Cref{sec:study}, we found that RoPE’s high-frequency components induce strong locality and drive reference copying (\Cref{fig:teaser}, \Cref{fig:copying}, \Cref{fig:method_abl}-b). A naïve attempt to mitigate this issue is to disable RoPE when attending to reference tokens; however, as shown in \Cref{fig:method_abl}-c, this leads to noticeable artifacts, indicating that RoPE’s positional structure remains necessary for coherent generation even under shared attention.

Motivated by this observation, we selectively modulate RoPE frequency components, attenuating high frequencies to reduce positional bias and amplifying low frequencies to promote global attention to the reference. Rather than splitting frequencies into discrete groups, we apply a frequency-aware modulation exclusively to the reference image keys and smoothly interpolate the modulation scale across the RoPE spectrum. Because the transition from low to high frequencies in RoPE is continuous, this design preserves stable attention behavior and avoids abrupt changes in positional sensitivity.
For smooth interpolation, we interpolate between a scale $s_{\text{hf}} \in (0,1)$ for the highest-frequency components and $s_{\text{lf}} > 1$ for the lowest-frequency components. We parameterize this interpolation using a polynomial schedule that increases smoothly from $s_{\text{hf}}$ to $s_{\text{lf}}$.
For each two-dimensional RoPE chunk $d$, we define a normalized index
$\tilde d = \frac{d}{\tfrac{D}{2}-1}$,
and assign a per-chunk scale:
\begin{align}
s_d = s_{\text{hf}} + \left(s_{\text{lf}} - s_{\text{hf}}\right)\, \tilde d^{\,\beta},
\qquad
d \in \left\{0, \dots, \tfrac{D}{2}-1 \right\}.
\label{eq:interp}
\end{align}
This formulation downscales high-frequency chunks (small $d$) toward $s_{\text{hf}}$, reducing positional bias, while low-frequency chunks (large $d$) approach $s_{\text{lf}}$, preserving global semantic guidance. We set $\beta = 2$ for all experiments, which we found to work best. We present an ablation on the value of $\beta$ in the supplementary.
We apply this modulation independently to the \(x\)- and \(y\)-axis partitions of the embeddings. Because, as noted in \Cref{subsec:rope}, the temporal partitions do not contribute to the positional sensitivity, they are always modulated by \(s_{\text{lf}}\).

In \Cref{fig:method_abl}-d, we further show that performing the reverse operation, scaling down the low-frequency chunks, does not mitigate the reference copying problem, which is consistent with our analysis in \Cref{sec:study}. 
In contrast, applying our approach with different modulation scales enables a controllable balance between content and style transfer. 
Using a larger scale for the high frequency chunks (\(s_{\text{hf}}\)) preserves the pose and overall structure of the reference while still transferring style and texture (\Cref{fig:method_abl}-e). 
Conversely, using a smaller scale for the high frequency chunks modifies the original structure while still achieving effective style transfer (\Cref{fig:method_abl}-f).

\paragraph{\textbf{Timestep scheduling}}
To account for the evolving behavior of the denoising process, where early timesteps establish global structure and later timesteps add texture and style~\cite{Issachar2025DyPE}, we linearly increase both \(s_{\text{hf}}\) and \(s_{\text{lf}}\) over time. 
\begin{wrapfigure}{r}{0.6\linewidth}
    \vspace{-5pt}
    \centering
    \begin{overpic}[width=1\linewidth]{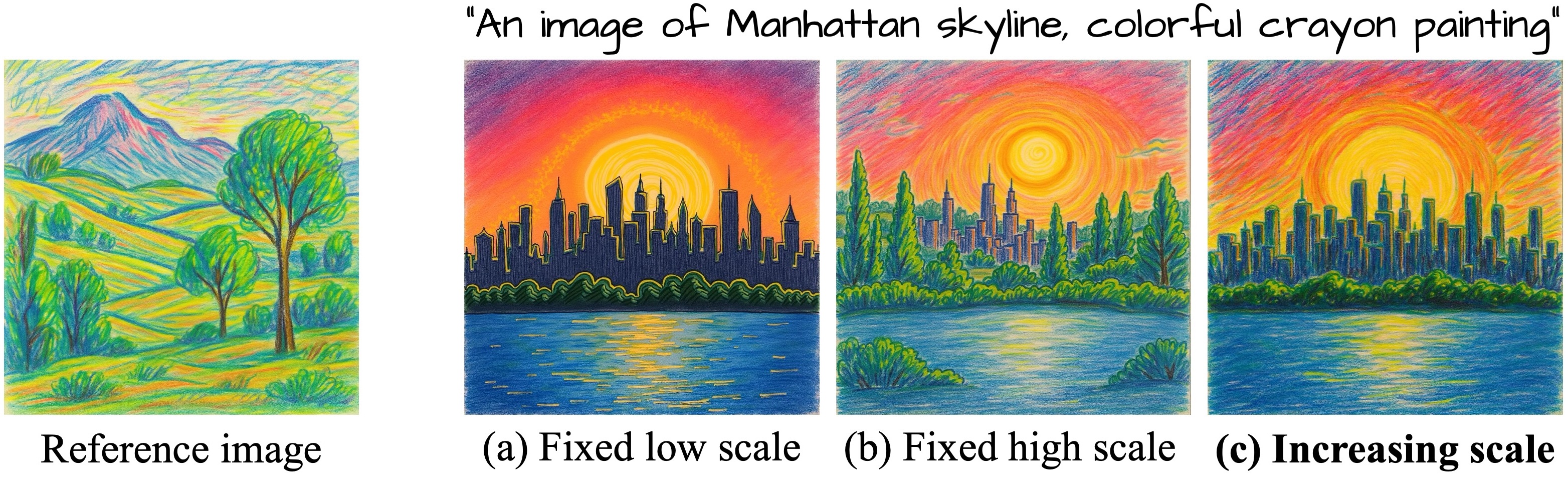}
    \end{overpic}
    \vspace{-20pt}
    \label{fig:timestep_ablation}
\end{wrapfigure}
Early in denoising, this reduces the influence of high frequency locality and encourages broad style and shape guidance. 
As denoising progresses, progressively increasing the scaling sharpens attention toward the reference, enabling accurate transfer of fine-grained stylistic details and textures. An example is shown in the inset figure. Using a fixed low
modulation scale for both $s_{hf}$ and $s_{lf}$ fails to transfer
fine stylistic attributes from the reference to the target
(inset-a).
Conversely, using a fixed high modulation
scale leads to excessive structural transfer from the reference
(inset-b). In contrast, our scheduling strategy allows
the model to first establish the correct global structure and subsequently,
at later denoising timesteps, increase attention to the reference to
transfer fine-grained details (inset-c).

With this modification, the key ($K$) for shared-attention in \Cref{eq:mmdit_sharing} becomes
\begin{align}
    K = 
    \bigl[ K^{\prime img}_{tar} \oplus K^{\prime txt}_{tar} \bigr]
    \,\oplus\,
    s^{FA}_{t} \cdot K^{img}_{ref},
    \quad
    s^{FA}_{t}=\{s_d^t\}_{d=0}^{\tfrac{D}{2}-1}, 
\end{align}
where \(s^{FA}_{t}\) is our frequency-aware modulation scale at denoising timestep \(t\). Only the reference image keys are modulated; target image keys, text keys, and all queries remain unchanged.

\paragraph{\textbf{Why Modulating RoPE’s Frequency Bands Works?}}

Our analysis in \Cref{sec:study} shows that RoPE decomposes attention into frequency components with distinct positional sensitivities: high frequency components impose strong locality, while low-frequency components support global interactions. This behavior is well-suited to standard image generation, where all tokens belong to a single evolving image and positional locality reinforces coherent local refinement without inducing competition.

In shared-attention settings, however, target and reference tokens share the same spatial grid, causing high-frequency RoPE components to dominate attention at aligned positions and induce reference copying. By modulating RoPE’s frequency bands, our method suppresses this positional dominance and shifts attention toward semantic similarity, enabling style transfer without content copying.

\section{Experiments}
\label{sec:experiments}
We evaluate our method on style-aligned generation and style transfer using \textbf{Flux 1.-dev}~\cite{flux2024} as the base model. 
Attention sharing is applied only to the single-stream blocks of the Flux DiT, which prior work~\cite{zhang2025alignedgen} shows primarily control appearance and style.
For style-aligned generation, we use the prompt set from StyleAligned~\cite{hertz2023StyleAligned}. For style transfer, reference images are taken from B-LoRA~\cite{frenkel2024implicit}, InstantStyle-Plus~\cite{wang2024instantstyleplus}, and other famous artworks. 


\paragraph{\textbf{Comparison}}
We compare our method against StyleAligned, which applies attention sharing in the UNet-based SDXL; AlignedGen, which performs attention sharing with shifted RoPE in Flux; IP-Adapter~\cite{flux-ipa}, trained for Flux by InstantX, which injects reference image features via an image encoder; and B-LoRA, which trains LoRA modules on a subset of SDXL blocks that are more sensitive to style.

We present a series of visual style transfer comparisons in \Cref{fig:visual_comparison}. While StyleAligned effectively captures the overall appearance of the reference images, it can introduce content leakage (\Cref{fig:summary}, \Cref{fig:visual_comparison}), and the transfer of stylistic attributes is often inconsistent. Moreover, relative to Flux, SDXL occasionally exhibits structural irregularities or repeated elements in the generated outputs, as shown in \Cref{fig:visual_comparison}.
AlignedGen generally produces coherent images; however, its use of shifted RoPE can induce content leakage by encouraging generation as a continuation of the reference. This results in unintended content transfer, such as the Tower of Pisa appearing in a beach scene (\Cref{fig:visual_comparison}, first row), or identity-related leakage across semantically distinct subjects (e.g., facial features transferring from the woman to the cat, \Cref{fig:visual_comparison}, fifth row).
IP-Adapter relies on a pretrained image encoder for reference feature extraction, which can limit its ability to generalize to reference images with unseen or out-of-distribution styles, and it may still exhibit content leakage similar to other methods. Likewise, B-LoRA struggles to consistently transfer style due to fine-tuning LoRA modules on a limited subset of SDXL blocks. In addition, B-LoRA requires per-reference training, making it more computationally demanding than our zero-shot approach.




\paragraph{\textbf{Discussion on shifted RoPE}}
A concurrent work, AlignedGen, mitigates reference copying by horizontally shifting the positional coordinates of reference tokens and applying RoPE to the reference keys with these shifted positions. This separation bypasses RoPE’s positional sensitivity and prevents direct copying. However, since target queries and keys remain positionally aligned, shifting the reference keys weakens attention to the reference, necessitating explicit modulation of the reference keys for faithful style transfer.

\begin{figure*}[t]
    \centering
    \begin{overpic}[width=1\linewidth]
    {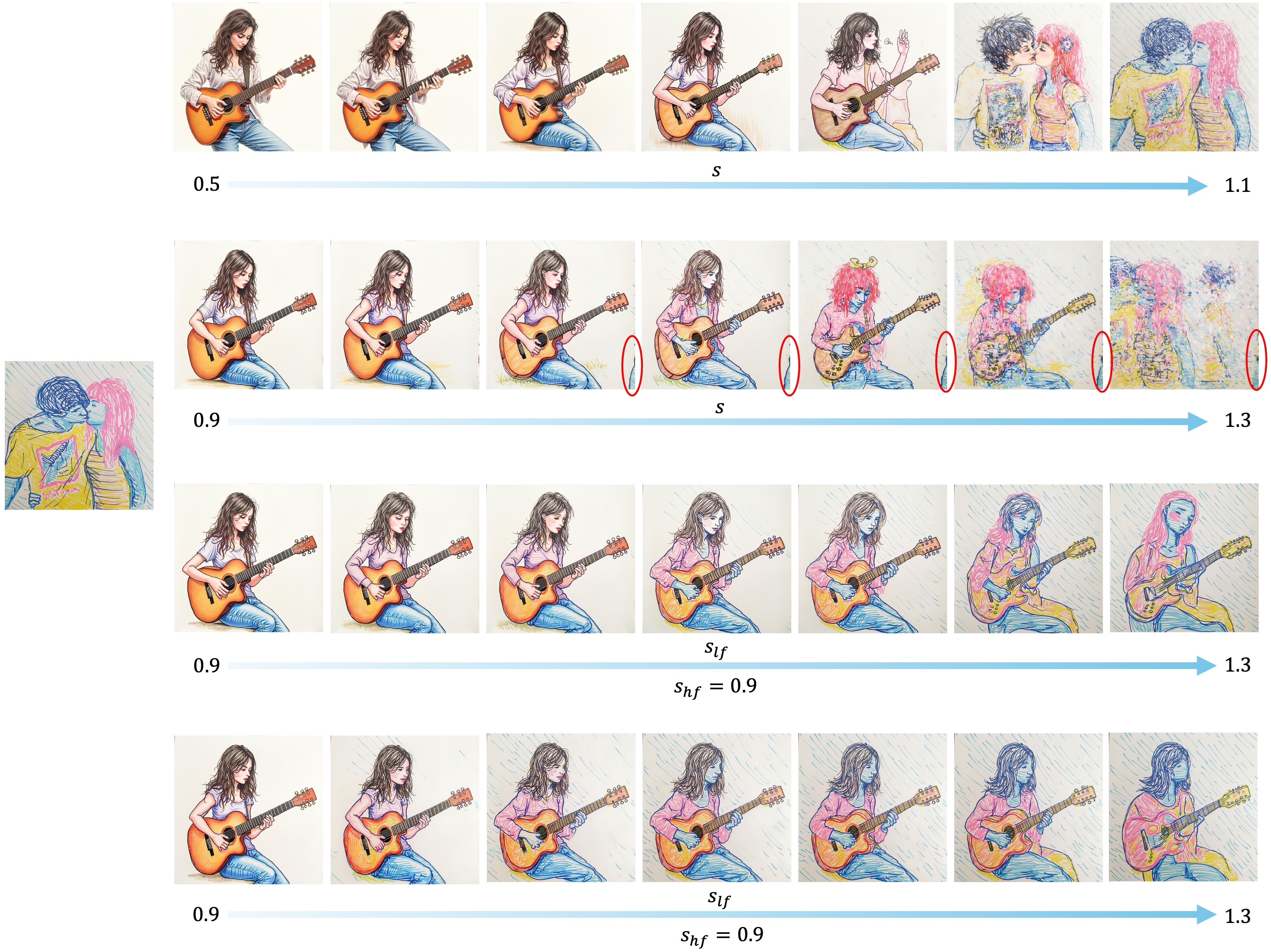}
    \put(-0.5, 47.5){Reference image}
    \put(44, 75.5){(a) Uniform modulation, w/o shift}
    \put(40, 57){(b) Uniform modulation, w/ shift (AlignedGen)}
    \put(39, 38){(c) Frequency-aware modulation, w/ shift}
    \put(36.5, 18.){(d) Frequency-aware modulation, w/o shift (\textbf{Ours})}
    \end{overpic}
    \caption{\textbf{Modulation and shifting effects.} 
    \textbf{(a)} Uniform modulation of reference tokens with a single scalar does not reliably transfer style; increasing the scale leads to content leakage and reference copying. 
    \textbf{(b)} Shifting reference token positions, as in AlignedGen~\cite{zhang2025alignedgen}, mitigates reference copying but introduces artifacts at larger scales due to the irregular behavior of high-frequency RoPE components. 
    \textbf{(c,d)} Our frequency-aware modulation addresses these issues both with and without positional shifting. 
    Without shifting, the overall attention to the reference is preserved, resulting in smoother and more controlled style transfer as $s_{lf}$ increases. The red circles show the ghost-cyclic inpainting problem of shifted RoPE.
}
    \label{fig:scale_ablation}
\end{figure*}

\Cref{fig:scale_ablation}-b shows that increasing the modulation scale $s$ initially improves style fidelity but eventually leads to structural incoherence and periodic reference copying due to high-frequency components of RoPE. Our frequency-based analysis resolves this issue even under shifted RoPE: as shown in \Cref{fig:scale_ablation}-c, suppressing the high-frequency components of the reference keys ($s_{hf}$) mitigates artifacts and copying, while amplifying the low-frequency components ($s_{lf}$) improves style fidelity.

As shown in \Cref{fig:scale_ablation}-d, when shifted RoPE positions are not used and attention to the reference is preserved, frequency-aware modulation yields the best results. In this setting, increasing $s_{lf}$ enables smooth and controlled style transfer. In contrast, as shown in \Cref{fig:scale_ablation}-a, uniform scaling without positional shifting is insufficient: small scales produce weak style transfer, while large scales cause content copying, consistent with \Cref{sec:study}.

\begin{figure*}[t]
    \centering
    \begin{overpic}[width=\linewidth]
    {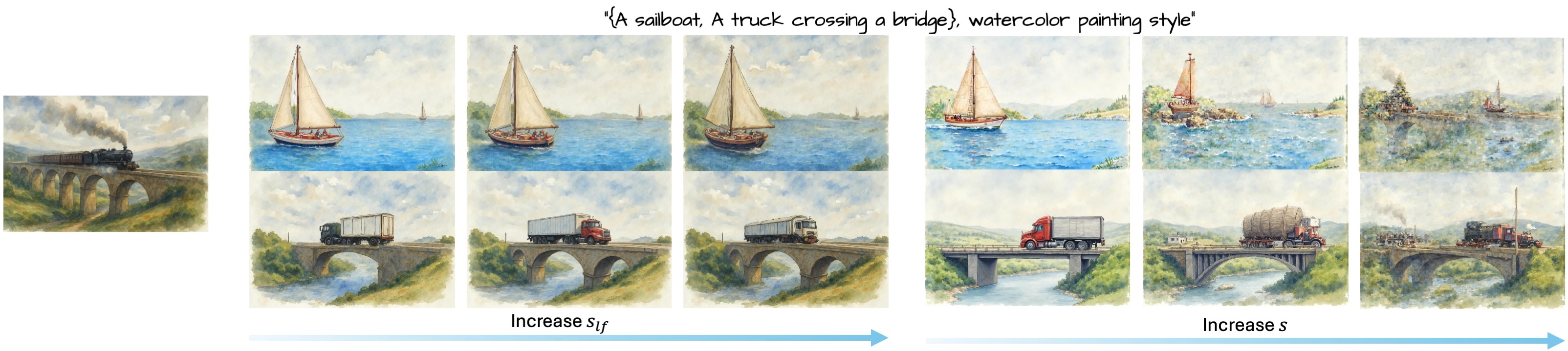}
    \end{overpic}
    \put(-335, 115){Ours}
    \put(-130, 115){Shifted RoPE}
    \caption{\textbf{High-resolution image generation.} In high-resolution settings (1536 $\times$ 1024), shifting reference tokens as in AlignedGen may introduce artifacts when reference attention increases due to the overly larger DiT context window consisting of unfamiliar RoPE positions. Our method avoids these artifacts and achieves better structural and stylistic alignment with the reference by increasing attention to reference tokens without positional shifts.
}

    \label{fig:wide_results}
\end{figure*}

As shown in \Cref{fig:wide_results}, the limitations of uniform modulation are amplified at high resolution. The large DiT context causes horizontally shifted reference tokens to have positional offsets unseen during training, leading to artifacts and periodic copying as the modulation scale increases. In contrast, our approach enables faithful style transfer, including shape and pose, by increasing $s_{lf}$.


A further limitation of shifted RoPE is that horizontally shifting the reference causes the model to interpret it as a continuation of the target image. When the reference contains incomplete structures, the model therefore completes them in the target, a phenomenon we term \emph{ghost-cyclic inpainting}. Examples are shown in \cref{fig:visual_comparison} and \Cref{fig:scale_ablation}-b, with the inpainted regions circled in red. Our frequency-aware modulation resolves this issue even under positional shifting, indicating that the effect stems from the locality induced by RoPE’s high-frequency components.







\section{Conclusion and future works}

We analyzed the role of rotary positional embeddings (RoPE) in diffusion transformers and showed that their frequency structure governs attention in shared-attention settings. High-frequency components enforce locality and lead to reference copying, while low-frequency components enable global, semantically driven interactions. Based on this insight, we introduced a frequency-aware modulation that enables controllable style-aligned generation without retraining or architectural changes.

Our findings highlight that positional encodings are not a neutral component, but actively govern the balance between locality and semantic association. This explains why attention-sharing strategies developed for UNet-based models do not directly transfer to DiTs and emphasizes the need to account for positional structure when designing shared-attention mechanisms in DiTs.

More broadly, this work suggests a direction for controlling generative models through targeted interventions in their internal representations. By analyzing how existing architectural components shape attention and information flow, future methods may achieve precise and interpretable control without introducing additional supervision or external control mechanisms.

\section*{Acknowledgements}
This research was supported by an NSERC Discovery Grant (RGPIN-2022-03111 and 2023-05617), the SFU Visual Computing Research Chair, and the Israel Science Foundation (grants no. 3441/21, 1473/24), Len Blavatnik, and the Blavatnik Family Foundation.
\clearpage


\bibliographystyle{ACM-Reference-Format}
\bibliography{main}

@String{Computing = "Computing" }

@String{Computer = "{IEEE} Computer" }

@String{Springer = "Springer-Verlag" }

@misc{labs2025flux1kontextflowmatching,
      title={FLUX.1 Kontext: Flow Matching for In-Context Image Generation and Editing in Latent Space},
      author={{Black Forest Labs}},
      year={2025},
      eprint={2506.15742},
      archivePrefix={arXiv},
      primaryClass={cs.GR},
      url={https://arxiv.org/abs/2506.15742},
}

@misc{flux2024,
    author={{Black Forest Labs}},
    title={FLUX},
    year={2024},
    howpublished={\url{https://github.com/black-forest-labs/flux}},
}

@article{Peebles2022DiT,
  title={Scalable Diffusion Models with Transformers},
  author={William Peebles and Saining Xie},
  year={2022},
  journal={arXiv preprint arXiv:2212.09748},
}

@misc{su2021roformer,
      title={RoFormer: Enhanced Transformer with Rotary Position Embedding}, 
      author={Jianlin Su and Yu Lu and Shengfeng Pan and Bo Wen and Yunfeng Liu},
      year={2021},
      eprint={2104.09864},
      archivePrefix={arXiv},
      primaryClass={cs.CL}
}

@article{hertz2023StyleAligned,
  title={Style Aligned Image Generation via Shared Attention},
  author={Hertz, Amir and Voynov, Andrey and Fruchter, Shlomi and Cohen-Or, Daniel},
  booktitle={arXiv preprint arxiv:2312.02133},
  year={2023}
}

@misc{tewel2024consistory,
        title={Training-Free Consistent Text-to-Image Generation}, 
        author={Yoad Tewel and Omri Kaduri and Rinon Gal and Yoni Kasten and Lior Wolf and Gal Chechik and Yuval Atzmon},
        year={2024},
        eprint={2402.03286},
        archivePrefix={arXiv},
        primaryClass={cs.CV},
        url={https://arxiv.org/abs/2402.03286}, 
}

@article{mou2023dragondiffusion,
  title={Dragondiffusion: Enabling drag-style manipulation on diffusion models},
  author={Mou, Chong and Wang, Xintao and Song, Jiechong and Shan, Ying and Zhang, Jian},
  journal={arXiv preprint arXiv:2307.02421},
  year={2023}
}

@inproceedings{huang2017adain,
  title={Arbitrary Style Transfer in Real-time with Adaptive Instance Normalization},
  author={Huang, Xun and Belongie, Serge},
  booktitle={ICCV},
  year={2017}
}

@article{zhang2025alignedgen,
  title={AlignedGen: Aligning Style Across Generated Images},
  author={Zhang, Jiexuan and Du, Yiheng and Wang, Qian and Li, Weiqi and Gu, Yu and Zhang, Jian},
  journal={arXiv preprint arXiv:2509.17088},
  year={2025}
}

@article{wei2025freeflux,
    title     = {FreeFlux: Understanding and Exploiting Layer-Specific Roles in RoPE-Based MMDiT for Versatile Image Editing},
    author    = {Wei, Tianyi and Zhou, yifan and Chen, Dongdong and Pan, Xingang},
    journal   = {Proceedings of the IEEE/CVF International Conference on Computer Vision},
    year      = {2025},
}

@InProceedings{Deng_2024_CVPR,
    author    = {Deng, Yingying and He, Xiangyu and Tang, Fan and Dong, Weiming},
    title     = {Z*: Zero-shot Style Transfer via Attention Reweighting},
    booktitle = {Proceedings of the IEEE/CVF Conference on Computer Vision and Pattern Recognition (CVPR)},
    month     = {June},
    year      = {2024},
    pages     = {6934-6944}
}

@article{deng2024z,
  title={Z-STAR+: A Zero-shot Style Transfer Method via Adjusting Style Distribution},
  author={Deng, Yingying and He, Xiangyu and Tang, Fan and Dong, Weiming},
  journal={arXiv preprint arXiv:2411.19231},
  year={2024}
}

@article{dosovitskiy2020vit,
  title={An Image is Worth 16x16 Words: Transformers for Image Recognition at Scale},
  author={Dosovitskiy, Alexey and Beyer, Lucas and Kolesnikov, Alexander and Weissenborn, Dirk and Zhai, Xiaohua and Unterthiner, Thomas and  Dehghani, Mostafa and Minderer, Matthias and Heigold, Georg and Gelly, Sylvain and Uszkoreit, Jakob and Houlsby, Neil},
  journal={ICLR},
  year={2021}
}

@inproceedings{vaswani2017attention,
author = {Vaswani, Ashish and Shazeer, Noam and Parmar, Niki and Uszkoreit, Jakob and Jones, Llion and Gomez, Aidan N. and Kaiser, \L{}ukasz and Polosukhin, Illia},
title = {Attention is all you need},
year = {2017},
isbn = {9781510860964},
publisher = {Curran Associates Inc.},
address = {Red Hook, NY, USA},
booktitle = {Proceedings of the 31st International Conference on Neural Information Processing Systems},
pages = {6000–6010},
numpages = {11},
location = {Long Beach, California, USA},
series = {NIPS'17}
}

@article{wan2025,
      title={Wan: Open and Advanced Large-Scale Video Generative Models}, 
      author={Team Wan and Ang Wang and Baole Ai and Bin Wen and Chaojie Mao and Chen-Wei Xie and Di Chen and Feiwu Yu and Haiming Zhao and Jianxiao Yang and Jianyuan Zeng and Jiayu Wang and Jingfeng Zhang and Jingren Zhou and Jinkai Wang and Jixuan Chen and Kai Zhu and Kang Zhao and Keyu Yan and Lianghua Huang and Mengyang Feng and Ningyi Zhang and Pandeng Li and Pingyu Wu and Ruihang Chu and Ruili Feng and Shiwei Zhang and Siyang Sun and Tao Fang and Tianxing Wang and Tianyi Gui and Tingyu Weng and Tong Shen and Wei Lin and Wei Wang and Wei Wang and Wenmeng Zhou and Wente Wang and Wenting Shen and Wenyuan Yu and Xianzhong Shi and Xiaoming Huang and Xin Xu and Yan Kou and Yangyu Lv and Yifei Li and Yijing Liu and Yiming Wang and Yingya Zhang and Yitong Huang and Yong Li and You Wu and Yu Liu and Yulin Pan and Yun Zheng and Yuntao Hong and Yupeng Shi and Yutong Feng and Zeyinzi Jiang and Zhen Han and Zhi-Fan Wu and Ziyu Liu},
      journal = {arXiv preprint arXiv:2503.20314},
      year={2025}
}

@inproceedings{Esser2024SD3,
  author    = {Patrick Esser and Sumith Kulal and Andreas Blattmann and others},
  title     = {Scaling Rectified Flow Transformers for High-Resolution Image Synthesis},
  booktitle = {Proceedings of the 41st International Conference on Machine Learning (ICML 2024)},
  year      = {2024},
  publisher = {JMLR.org},
  doi       = {10.5555/3692070.3692573}
}

@inproceedings{Podell2024SDXL,
  author    = {Dustin Podell and Zion English and Kyle Lacey and Andreas Blattmann and Tim Dockhorn and Jonas Müller and Joe Penna and Robin Rombach},
  title     = {SDXL: Improving Latent Diffusion Models for High-Resolution Image Synthesis},
  booktitle = {Proceedings of the International Conference on Learning Representations (ICLR) 2024},
  year      = {2024},
  note      = {Spotlight presentation},
  url       = {https://openreview.net/forum?id=di52zR8xgf}
}

@inproceedings{Rombach2022LDM,
  author    = {Rombach, Robin and Blattmann, Andreas and Lorenz, Dominik and Esser, Patrick and Ommer, Bj{\"o}rn},
  title     = {High-Resolution Image Synthesis With Latent Diffusion Models},
  booktitle = {Proceedings of the IEEE/CVF Conference on Computer Vision and Pattern Recognition (CVPR)},
  year      = {2022},
  pages     = {10684--10695},
  url       = {https://openaccess.thecvf.com/content/CVPR2022/papers/Rombach_High-Resolution_Image_Synthesis_With_Latent_Diffusion_Models_CVPR_2022_paper.pdf}
}

@article{Issachar2025DyPE,
  title   = {DyPE: Dynamic Position Extrapolation for Ultra High Resolution Diffusion},
  author  = {Noam Issachar and Guy Yariv and Sagie Benaim and Yossi Adi and Dani Lischinski and Raanan Fattal},
  journal = {arXiv preprint arXiv:2510.20766},
  year    = {2025},
  url     = {https://arxiv.org/abs/2510.20766}
}

@article{Chen2023PositionalInterpolation,
  title   = {Extending Context Window of Large Language Models via Positional Interpolation},
  author  = {Shouyuan Chen and Sherman Wong and Liangjian Chen and Yuandong Tian},
  journal = {arXiv preprint arXiv:2306.15595},
  year    = {2023},
  url     = {https://arxiv.org/abs/2306.15595}
}

@misc{Reddit2023NTKAwareRoPE,
  title        = {NTK-Aware Scaled RoPE Allows LLaMA Models to Have Larger Context Windows},
  howpublished = {\url{https://www.reddit.com/r/LocalLLaMA/comments/14lz7j5/ntkaware_scaled_rope_allows_llama_models_to_have/}},
  note         = {Reddit discussion post},
  year         = {2023},
  month        = {June}
}

@inproceedings{Peng2024YaRN,
  title     = {YaRN: Efficient Context Window Extension of Large Language Models},
  author    = {Bowen Peng and Jeffrey Quesnelle and Honglu Fan and Enrico Shippole},
  booktitle = {The Twelfth International Conference on Learning Representations (ICLR) 2024},
  year      = {2024},
  url       = {https://openreview.net/forum?id=wHBfxhZu1u}
}

@InProceedings{Avrahami_2025_CVPR,
    author    = {Avrahami, Omri and Patashnik, Or and Fried, Ohad and Nemchinov, Egor and Aberman, Kfir and Lischinski, Dani and Cohen-Or, Daniel},
    title     = {Stable Flow: Vital Layers for Training-Free Image Editing},
    booktitle = {Proceedings of the Computer Vision and Pattern Recognition Conference (CVPR)},
    month     = {June},
    year      = {2025},
    pages     = {7877-7888}
}

@article{wang2024taming,
  title={Taming Rectified Flow for Inversion and Editing},
  author={Wang, Jiangshan and Pu, Junfu and Qi, Zhongang and Guo, Jiayi and Ma, Yue and Huang, Nisha and Chen, Yuxin and Li, Xiu and Shan, Ying},
  journal={arXiv preprint arXiv:2411.04746},
  year={2024}
}

@InProceedings{cao_2023_masactrl,
    author    = {Cao, Mingdeng and Wang, Xintao and Qi, Zhongang and Shan, Ying and Qie, Xiaohu and Zheng, Yinqiang},
    title     = {MasaCtrl: Tuning-Free Mutual Self-Attention Control for Consistent Image Synthesis and Editing},
    booktitle = {Proceedings of the IEEE/CVF International Conference on Computer Vision (ICCV)},
    month     = {October},
    year      = {2023},
    pages     = {22560-22570}
}

@InProceedings{Tumanyan_2023_CVPR,
        author    = {Tumanyan, Narek and Geyer, Michal and Bagon, Shai and Dekel, Tali},
        title     = {Plug-and-Play Diffusion Features for Text-Driven Image-to-Image Translation},
        booktitle = {Proceedings of the IEEE/CVF Conference on Computer Vision and Pattern Recognition (CVPR)},
        month     = {June},
        year      = {2023},
        pages     = {1921-1930}
    }

@misc{alaluf2023crossimage,
        title={Cross-Image Attention for Zero-Shot Appearance Transfer}, 
        author={Yuval Alaluf and Daniel Garibi and Or Patashnik and Hadar Averbuch-Elor and Daniel Cohen-Or},
        year={2023},
        eprint={2311.03335},
        archivePrefix={arXiv},
        primaryClass={cs.CV}
      }

@misc{almohammadi2025coracorrespondenceawareimageediting,
      title={Cora: Correspondence-aware image editing using few step diffusion}, 
      author={Amirhossein Alimohammadi and Aryan Mikaeili and Sauradip Nag and Negar Hassanpour and Andrea Tagliasacchi and Ali Mahdavi-Amiri},
      year={2025},
      eprint={2505.23907},
      archivePrefix={arXiv},
      primaryClass={cs.CV},
      url={https://arxiv.org/abs/2505.23907}, 
}

@article{koo2024flexiedit,
    title={FlexiEdit: Frequency-Aware Latent Refinement for Enhanced Non-Rigid Editing},
    author={Koo, Gwanhyeong and Yoon, Sunjae and Hong, Ji Woo and Yoo, Chang D},
    journal={arXiv preprint arXiv:2407.17850},
    year={2024}}

@inproceedings{patashnik2024qnerf,
author = {Patashnik, Or and Gal, Rinon and Cohen-Or, Daniel and Zhu, Jun-Yan and De La Torre, Fernando},
title = {Consolidating Attention Features for Multi-view Image Editing},
year = {2024},
isbn = {9798400711312},
publisher = {Association for Computing Machinery},
address = {New York, NY, USA},
url = {https://doi.org/10.1145/3680528.3687611},
doi = {10.1145/3680528.3687611},
booktitle = {SIGGRAPH Asia 2024 Conference Papers},
articleno = {40},
numpages = {12},
location = {Tokyo, Japan},
series = {SA '24}
}

@article{xu2023infedit,
  title={Inversion-Free Image Editing with Natural Language}, 
  author={Sihan Xu and Yidong Huang and Jiayi Pan and Ziqiao Ma and Joyce Chai},
  booktitle={Conference on Computer Vision and Pattern Recognition 2024},
  year={2024}
}

@misc{mikaeili2025griffingenerativereferencelayout,
      title={Griffin: Generative Reference and Layout Guided Image Composition}, 
      author={Aryan Mikaeili and Amirhossein Alimohammadi and Negar Hassanpour and Ali Mahdavi-Amiri and Andrea Tagliasacchi},
      year={2025},
      eprint={2509.23643},
      archivePrefix={arXiv},
      primaryClass={cs.CV},
      url={https://arxiv.org/abs/2509.23643}, 
}

@InProceedings{generativephotomontage,
  author = {Liu, Sean J. and Kumari, Nupur and Shamir, Ariel and Zhu, Jun-Yan},
  title = {Generative Photomontage},
  booktitle = {Proceedings of the Computer Vision and Pattern Recognition Conference (CVPR)},
  month = {June},
  year = {2025},
  pages = {7931-7941},
}

@inproceedings{
eldesokey2025buildascene,
title={Build-A-Scene: Interactive 3D Layout Control for Diffusion-Based Image Generation},
author={Abdelrahman Eldesokey and Peter Wonka},
booktitle={The Thirteenth International Conference on Learning Representations},
year={2025},
url={https://openreview.net/forum?id=gg6dPtdC1C}
}

@article{tokenflow2023,
        title = {TokenFlow: Consistent Diffusion Features for Consistent Video Editing},
        author = {Geyer, Michal and Bar-Tal, Omer and Bagon, Shai and Dekel, Tali},
        journal={arXiv preprint arxiv:2307.10373},
        year={2023}
        }

@article{qi2023fatezero,
      title={FateZero: Fusing Attentions for Zero-shot Text-based Video Editing}, 
      author={Chenyang Qi and Xiaodong Cun and Yong Zhang and Chenyang Lei and Xintao Wang and Ying Shan and Qifeng Chen},
      year={2023},
      journal={arXiv:2303.09535},
}

@misc{wu2025qwenimagetechnicalreport,
      title={Qwen-Image Technical Report}, 
      author={Chenfei Wu and Jiahao Li and Jingren Zhou and Junyang Lin and Kaiyuan Gao and Kun Yan and Sheng-ming Yin and Shuai Bai and Xiao Xu and Yilei Chen and Yuxiang Chen and Zecheng Tang and Zekai Zhang and Zhengyi Wang and An Yang and Bowen Yu and Chen Cheng and Dayiheng Liu and Deqing Li and Hang Zhang and Hao Meng and Hu Wei and Jingyuan Ni and Kai Chen and Kuan Cao and Liang Peng and Lin Qu and Minggang Wu and Peng Wang and Shuting Yu and Tingkun Wen and Wensen Feng and Xiaoxiao Xu and Yi Wang and Yichang Zhang and Yongqiang Zhu and Yujia Wu and Yuxuan Cai and Zenan Liu},
      year={2025},
      eprint={2508.02324},
      archivePrefix={arXiv},
      primaryClass={cs.CV},
      url={https://arxiv.org/abs/2508.02324}, 
}

@inbook{efros2023quilting,
author = {Efros, Alexei A. and Freeman, William T.},
title = {Image Quilting for Texture Synthesis and Transfer},
year = {2023},
isbn = {9798400708978},
publisher = {Association for Computing Machinery},
address = {New York, NY, USA},
edition = {1},
url = {https://doi.org/10.1145/3596711.3596771},
booktitle = {Seminal Graphics Papers: Pushing the Boundaries, Volume 2},
articleno = {59},
numpages = {6}
}

@inproceedings{hertzmann2001analogies,
author = {Hertzmann, Aaron and Jacobs, Charles E. and Oliver, Nuria and Curless, Brian and Salesin, David H.},
title = {Image analogies},
year = {2001},
isbn = {158113374X},
publisher = {Association for Computing Machinery},
address = {New York, NY, USA},
url = {https://doi.org/10.1145/383259.383295},
doi = {10.1145/383259.383295},
booktitle = {Proceedings of the 28th Annual Conference on Computer Graphics and Interactive Techniques},
pages = {327–340},
numpages = {14},
series = {SIGGRAPH '01}
}

@inproceedings{Gatys2016StyleTransfer,
  author    = {Gatys, Leon A. and Ecker, Alexander S. and Bethge, Matthias},
  title     = {Image Style Transfer Using Convolutional Neural Networks},
  booktitle = {Proceedings of the IEEE Conference on Computer Vision and Pattern Recognition (CVPR)},
  year      = {2016},
  pages     = {2414--2423},
  url       = {https://www.cv-foundation.org/openaccess/content_cvpr_2016/papers/Gatys_Image_Style_Transfer_CVPR_2016_paper.pdf}
}

@inproceedings{Johnson2016PerceptualLosses,
  author    = {Johnson, Justin and Alahi, Alexandre and Fei-Fei, Li},
  title     = {Perceptual Losses for Real-Time Style Transfer and Super-Resolution},
  booktitle = {Proceedings of the European Conference on Computer Vision (ECCV)},
  year      = {2016},
  pages     = {694--711},
  publisher = {Springer},
  url       = {https://cs.stanford.edu/people/jcjohns/papers/eccv16/JohnsonECCV16.pdf}
}

@inproceedings{CycleGAN2017,
  title={Unpaired Image-to-Image Translation using Cycle-Consistent Adversarial Networks},
  author={Zhu, Jun-Yan and Park, Taesung and Isola, Phillip and Efros, Alexei A},
  booktitle={Computer Vision (ICCV), 2017 IEEE International Conference on},
  year={2017}
}

@inproceedings{huang2018munit,
  title={Multimodal Unsupervised Image-to-image Translation},
  author={Huang, Xun and Liu, Ming-Yu and Belongie, Serge and Kautz, Jan},
  booktitle={ECCV},
  year={2018}
}

@article{pix2pix2017,
  title={Image-to-Image Translation with Conditional Adversarial Networks},
  author={Isola, Phillip and Zhu, Jun-Yan and Zhou, Tinghui and Efros, Alexei A},
  journal={CVPR},
  year={2017}
}

@misc{gal2022textual,
      doi = {10.48550/ARXIV.2208.01618},
      url = {https://arxiv.org/abs/2208.01618},
      author = {Gal, Rinon and Alaluf, Yuval and Atzmon, Yuval and Patashnik, Or and Bermano, Amit H. and Chechik, Gal and Cohen-Or, Daniel},
      title = {An Image is Worth One Word: Personalizing Text-to-Image Generation using Textual Inversion},
      publisher = {arXiv},
      year = {2022},
      primaryClass={cs.CV}
}

@article{ruiz2022dreambooth,
  title={DreamBooth: Fine Tuning Text-to-image Diffusion Models for Subject-Driven Generation},
  author={Ruiz, Nataniel and Li, Yuanzhen and Jampani, Varun and Pritch, Yael and Rubinstein, Michael and Aberman, Kfir},
  booktitle={arXiv preprint arxiv:2208.12242},
  year={2022}
}

@misc{frenkel2024implicit,
      title={Implicit Style-Content Separation using B-LoRA}, 
      author={Yarden Frenkel and Yael Vinker and Ariel Shamir and Daniel Cohen-Or},
      year={2024},
      eprint={2403.14572},
      archivePrefix={arXiv},
      primaryClass={cs.CV}
}

@inproceedings{Sohn2023StyleDrop,
  author    = {Kihyuk Sohn and Lu Jiang and Jarred Barber and Kimin Lee and Nataniel Ruiz and Dilip Krishnan and Huiwen Chang and Yuanzhen Li and Irfan Essa and Michael Rubinstein and Yuan Hao and Glenn Entis and Irina Blok and Daniel Castro Chin},
  title     = {StyleDrop: Text-to-Image Synthesis of Any Style},
  booktitle = {Advances in Neural Information Processing Systems (NeurIPS) 2023},  
  year      = {2023},
  note      = {Poster — OpenReview preprint},
  url       = {https://openreview.net/forum?id=KoaFh16uOc}
}

@article{ye2023ip-adapter,
  title={IP-Adapter: Text Compatible Image Prompt Adapter for Text-to-Image Diffusion Models},
  author={Ye, Hu and Zhang, Jun and Liu, Sibo and Han, Xiao and Yang, Wei},
  booktitle={arXiv preprint arxiv:2308.06721},
  year={2023}
}

@article{wang2024instantstyle,
  title={Instantstyle: Free lunch towards style-preserving in text-to-image generation},
  author={Wang, Haofan and Wang, Qixun and Bai, Xu and Qin, Zekui and Chen, Anthony},
  journal={arXiv preprint arXiv:2404.02733},
  year={2024}
}

@article{wang2024instantstyleplus,
      title={InstantStyle-Plus: Style Transfer with Content-Preserving in Text-to-Image Generation},
      author={Wang, Haofan and Xing, Peng and Huang, Renyuan and Ai, Hao and Wang, Qixun and Bai, Xu},
      journal={arXiv preprint arXiv:2407.00788},
      year={2024}
    }

@article{liu2023stylecrafter,
  title={StyleCrafter: Enhancing Stylized Text-to-Video Generation with Style Adapter},
  author={Liu, Gongye and Xia, Menghan and Zhang, Yong and Chen, Haoxin and Xing, Jinbo and Wang, Xintao and Yang, Yujiu and Shan, Ying},
  journal={arXiv preprint arXiv:2312.00330},
  year={2023}
}

@misc{chen2025devil,
  title={The Devil is in Attention Sharing: Improving Complex Non-rigid Image Editing Faithfulness via Attention Synergy},
  author={Zhuo Chen and Fanyue Wei and Runze Xu and Jingjing Li and Lixin Duan and Angela Yao and Wen Li},
  year={2025},
  eprint={2512.14423},
  archivePrefix={arXiv},
  primaryClass={cs.CV},
  url={https://arxiv.org/abs/2512.14423},
}

@inproceedings{ye2025stylemaster,
  title={Stylemaster: Stylize your video with artistic generation and translation},
  author={Ye, Zixuan and Huang, Huijuan and Wang, Xintao and Wan, Pengfei and Zhang, Di and Luo, Wenhan},
  booktitle={Proceedings of the Computer Vision and Pattern Recognition Conference},
  pages={2630--2640},
  year={2025}
}

@article{bai2025positional,
      title={Positional Encoding Field},
      author={Bai, Yunpeng and Li, Haoxiang and Huang, Qixing},
      journal={arXiv preprint arXiv:2510.20385},
      year={2025}
    }

@article{bahmani2025ac3d,
  author = {Bahmani, Sherwin and Skorokhodov, Ivan and Qian, Guocheng and Siarohin, Aliaksandr and Menapace, Willi and Tagliasacchi, Andrea and Lindell, David B. and Tulyakov, Sergey},
  title = {AC3D: Analyzing and Improving 3D Camera Control in Video Diffusion Transformers},
  journal = {Proc. CVPR},
  year = {2025},
}

@misc{flux-ipa,
    author = {{InstantX Team}},
    title = {InstantX FLUX.1-dev IP-Adapter Page},
    year = {2024},
}

\begin{figure*}[t]
    \centering
    \begin{overpic}[width=\linewidth]
    {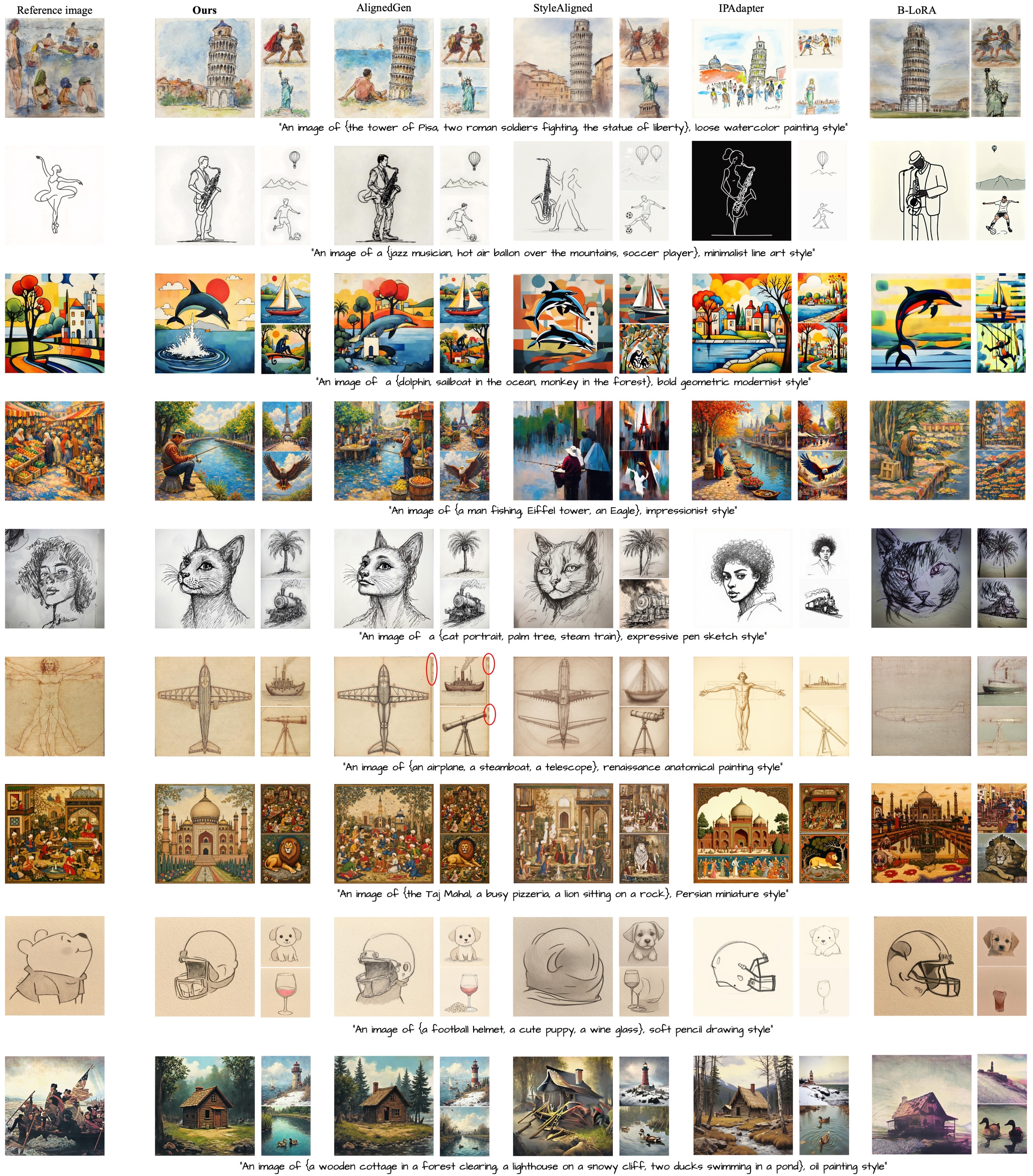}
    \end{overpic}
    \caption{\textbf{Comparison.} 
We visually compare our method’s style transfer results with AlignedGen, StyleAligned, IPAdapter, and B-LoRA. 
Our approach achieves faithful style transfer while maintaining structural coherence, avoiding (irrelevant to the target prompt) content leakage from the reference to the generated image. The red circles show the ghost-cyclic inpainting problem of shifted RoPE.
}
   \label{fig:visual_comparison}
\end{figure*}


\clearpage
\clearpage
\setcounter{page}{1}
\appendix

\section{Scaling interpolation function}
As described in \Cref{eq:interp}, we employ a polynomial function to schedule the interpolation scale. While all experiments in the main paper use $\beta = 2$, here we analyze the effect of this hyperparameter.

\begin{figure}[!h]
    \centering
    \begin{overpic}[width=0.8\linewidth]{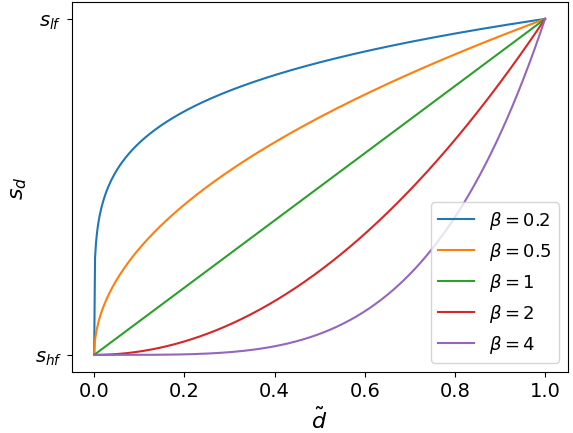}
    \end{overpic}
    \caption{Polynomial scheduling curves for different values of $\beta$ (all start at $s_{\text{hf}}$ and end at $s_{\text{lf}}$).}
    \label{fig:poly_plot}
\end{figure}

\Cref{fig:poly_plot} illustrates the interpolation function for different values of $\beta$. Although all curves start at $s_{\text{hf}}$ and end at $s_{\text{lf}}$, they differ in their intermediate behavior, which controls how quickly the transition occurs between the two endpoints.

\FloatBarrier 

\begin{figure}[!h]
    \centering
    \begin{overpic}[width=0.8\linewidth]{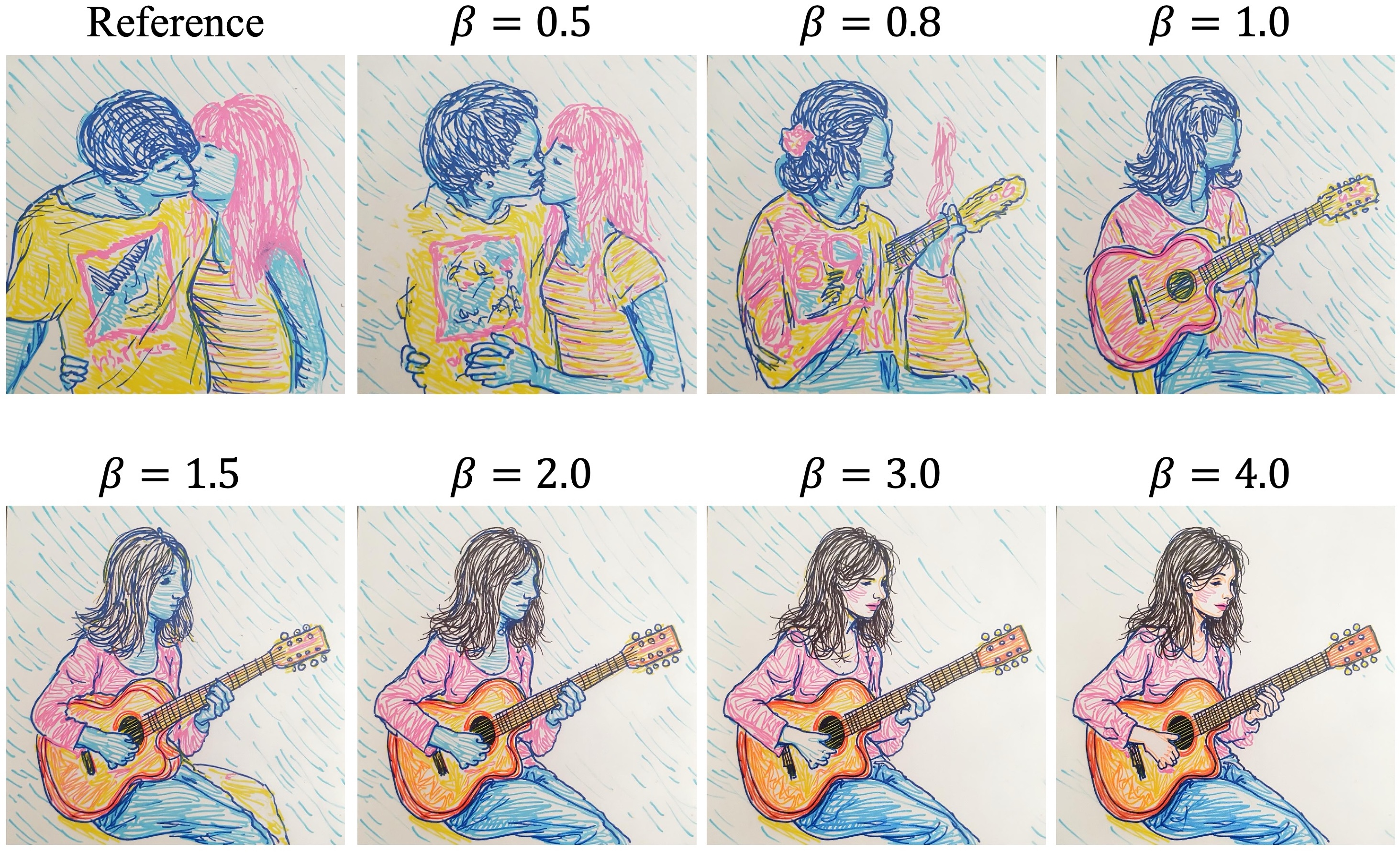}
    \end{overpic}
    \caption{\textbf{Qualitative ablation of $\beta$ on style transfer behavior.}}
    \label{fig:beta_ablation}
\end{figure}

\FloatBarrier 

\section{User study}
To compare our method with the baselines described in \Cref{sec:experiments}, we conduct a user study. The study consists of 20 style transfer examples, with reference images drawn from the B-LoRA dataset, the InstantStyle-Plus dataset, and other well-known artworks. Participants are asked to select and rank the top three results among five methods based on the following criteria: \textbf{(1)} style alignment, \textbf{(2)} alignment with the target prompt without unwanted content leakage from the reference, and \textbf{(3)} structural correctness (e.g., no missing or duplicated body parts such as extra hands).

For scoring, the first-ranked result receives 3 points, the second receives 2 points, the third receives 1 point, and the remaining two receive 0 points. \Cref{tab:user_study_comp} reports the aggregated results over 25 participants, showing that our method outperforms all baselines.

\begin{table}[h]
    \centering
    \captionsetup{font=small}
    \caption{\textbf{User study}. Our method receives a significantly higher score than the alternatives.}
    \small
    \setlength{\tabcolsep}{3.5pt} 

    \resizebox{\linewidth}{!}{%
    \begin{tabular}{c|ccccc}
        Method &
        B-LoRA &
        IPAdapter &
        StyleAligned &
        AlignedGen &
        \textbf{Ours} \\
        \midrule
        Average Ranking ($\uparrow$) &
        1.15 &
        0.44 &
        0.74 &
        1.26 &
        \textbf{2.40} \\
    \end{tabular}%
    }
    \label{tab:user_study_comp}
\end{table}

\FloatBarrier

\section{Comparison with Flux Kontext}
\begin{figure*}[t]
    \centering
    \begin{overpic}[width=0.9\linewidth]
    {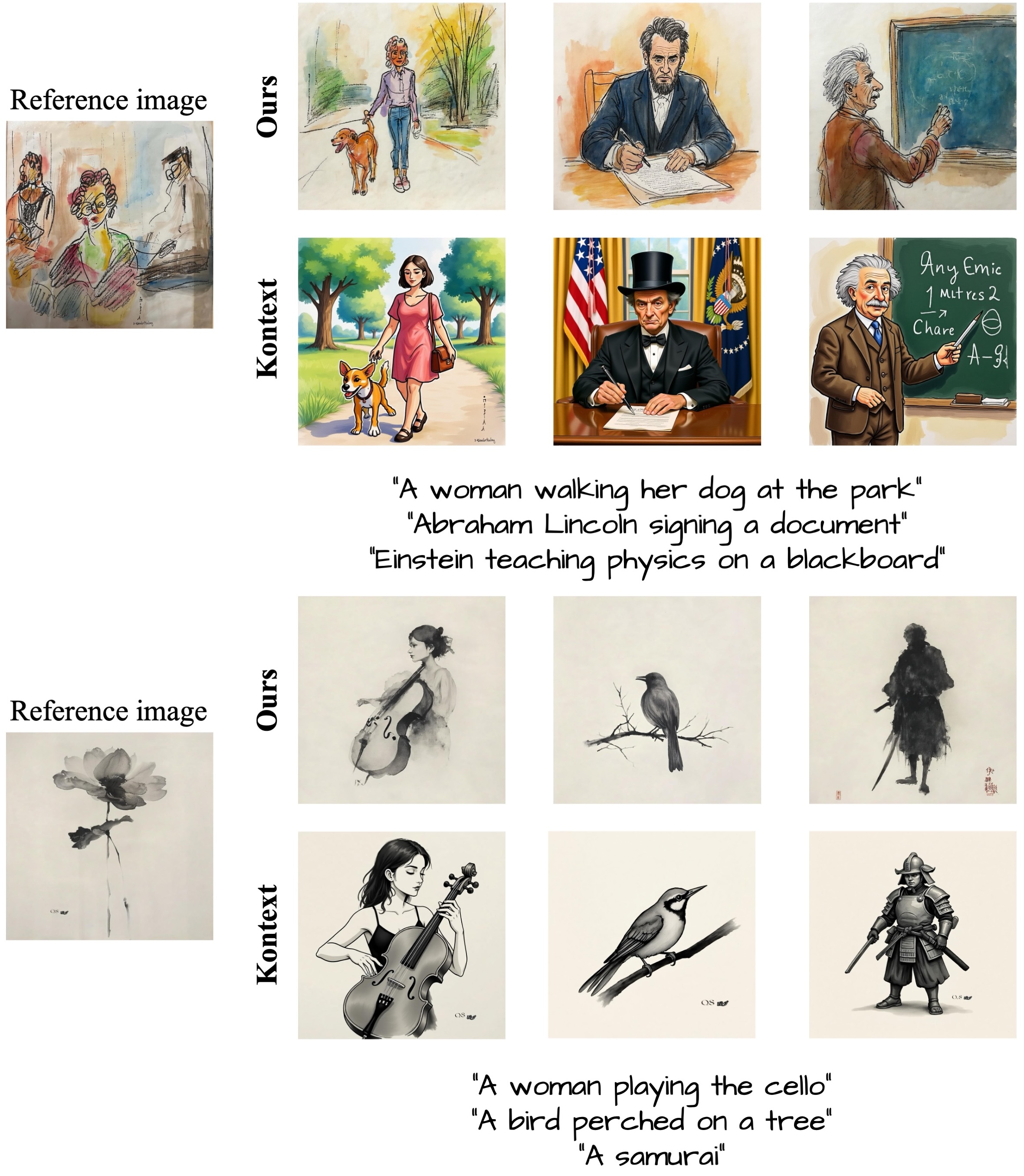}
    \end{overpic}
    \caption{\textbf{Comparison with Flux Kontext.} Compared to Flux Kontext, our method is able to transfer finegrained stylistic cues from the reference image.
}
\label{fig:kontext}
\end{figure*}

In this section, we compare our method with Flux Kontext~\cite{labs2025flux1kontextflowmatching}, a fine-tuned variant of FLUX designed for in-context generation and image conditioning. The results in \Cref{fig:kontext} show that our method significantly outperforms Kontext. While Kontext performs reasonably well for coarse styles (bottom example), it struggles to capture fine-grained stylistic details, shapes, and brush strokes (top example). This limitation is likely due to Kontext being primarily trained for image editing rather than style transfer.

\section{Extra results}
\begin{figure}[!h]
    \centering
    \begin{overpic}[width=0.9\linewidth]{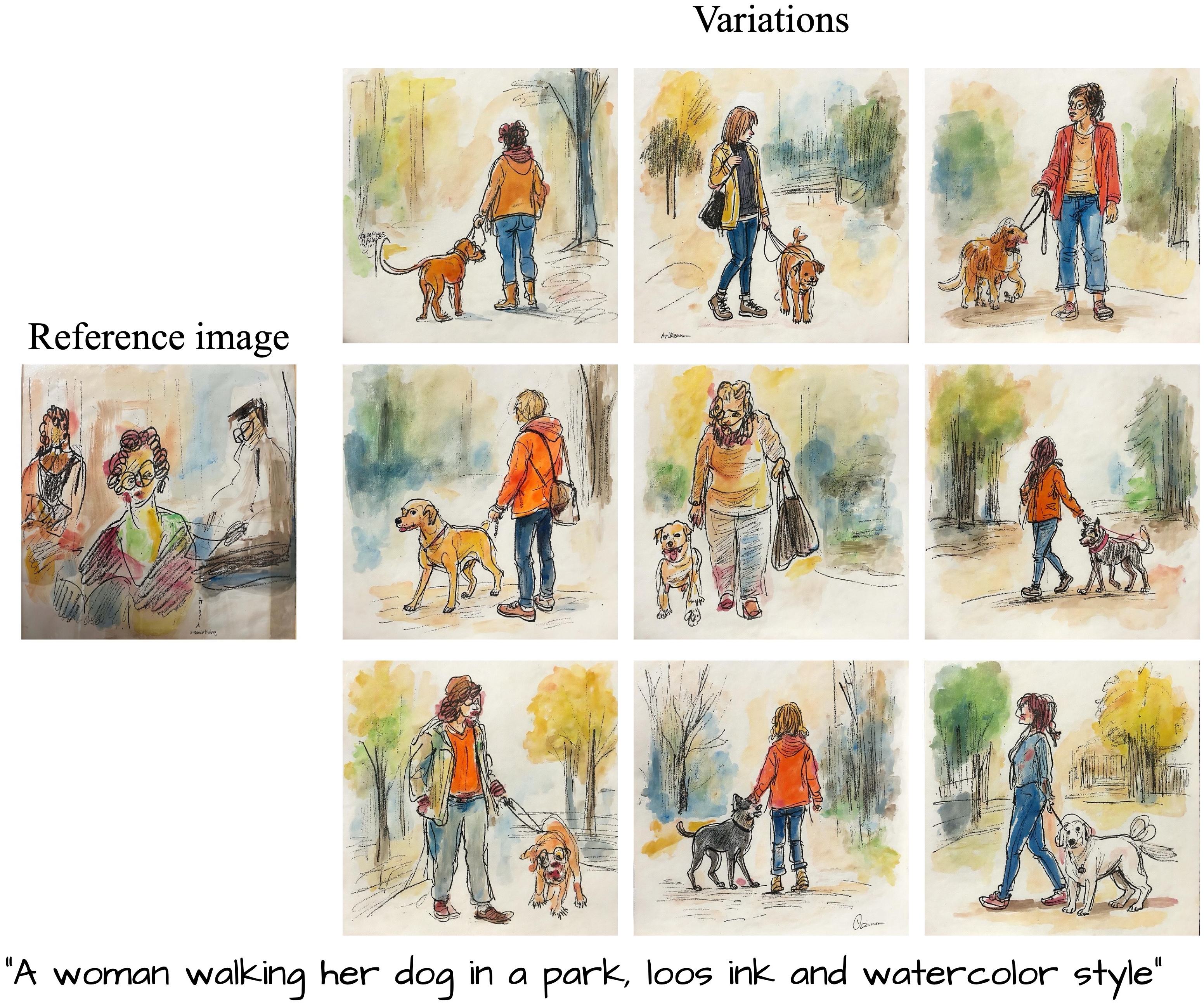}
    \end{overpic}
    \caption{\textbf{Generation results from different initial noise.} We are able to generate variations that respect the style of the reference image and the text prompt.}
    \label{fig:var}
\end{figure}

\Cref{fig:var} demonstrates that, from different noise maps, our method can generate diverse outputs from the same text prompt while consistently preserving the style of the reference image.

\begin{figure*}[t]
    \centering
    \begin{overpic}[width=0.9\linewidth]
    {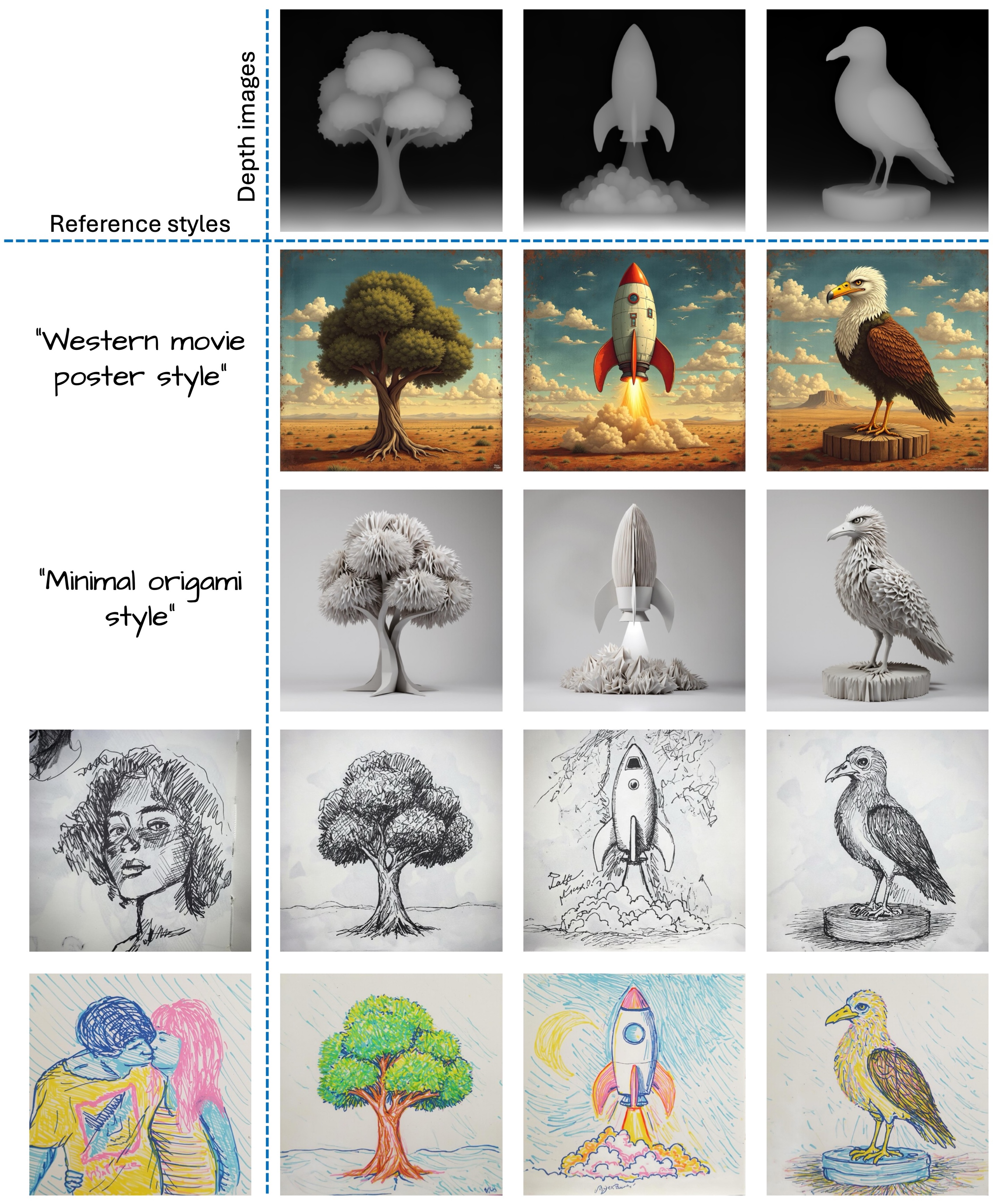}
    \end{overpic}
    \caption{\textbf{Depth conditioned generation.} Given an input depth map and a reference style, our method generates images that preserve the depth structure while faithfully transferring diverse styles across different objects.
}
\label{fig:depth}
\end{figure*}

In \Cref{fig:depth} we show results of our method on FLUX.1-Depth-dev, which is fine-tuned for depth-conditioned image generation.

\begin{figure*}[t]
    \centering
    \begin{overpic}[width=0.7\linewidth]
    {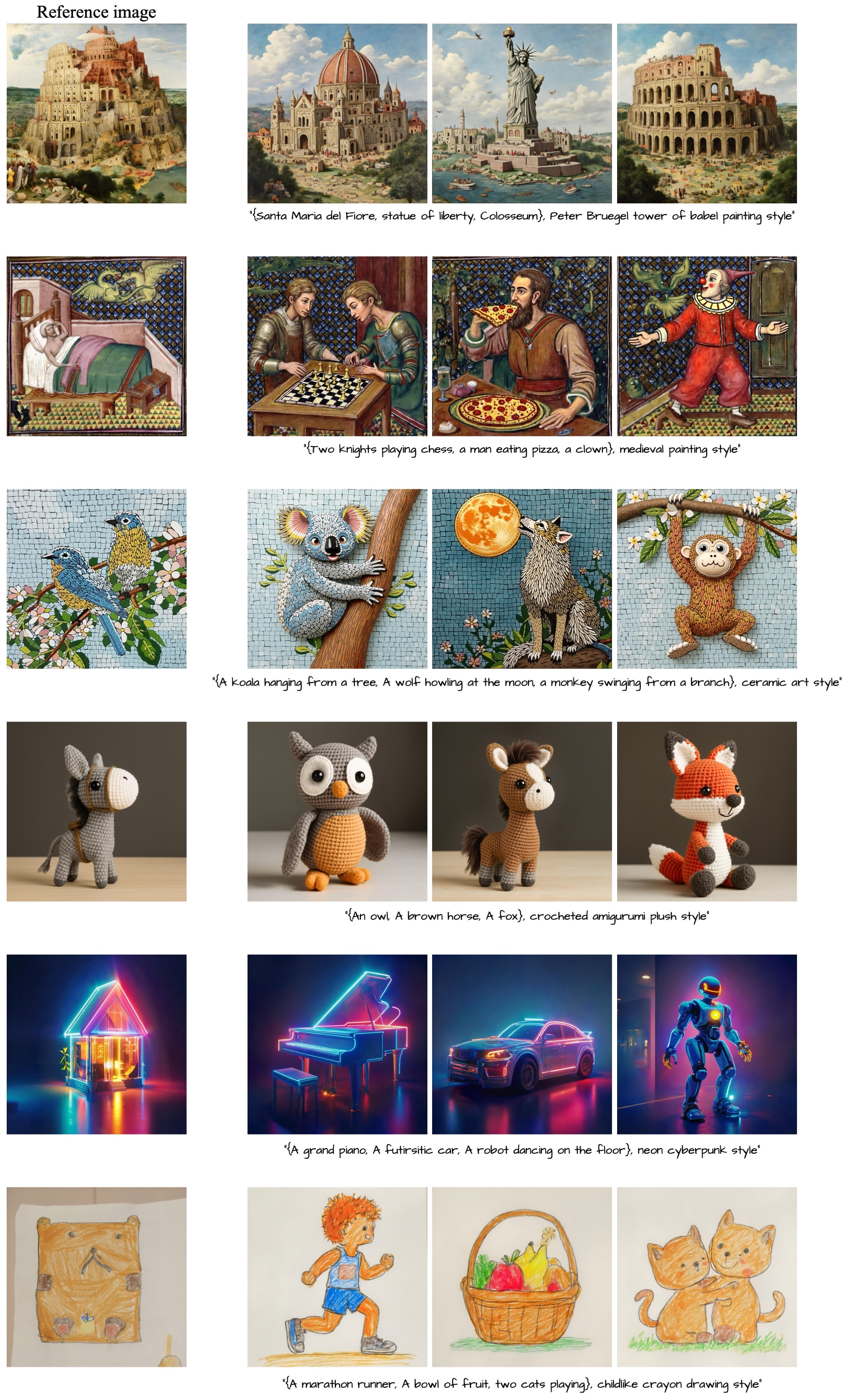}
    \end{overpic}
    \caption{\textbf{Extra style transfer results}
}
\label{fig:extra_1}
\end{figure*}

\begin{figure*}[t]
    \centering
    \begin{overpic}[width=0.65\linewidth]
    {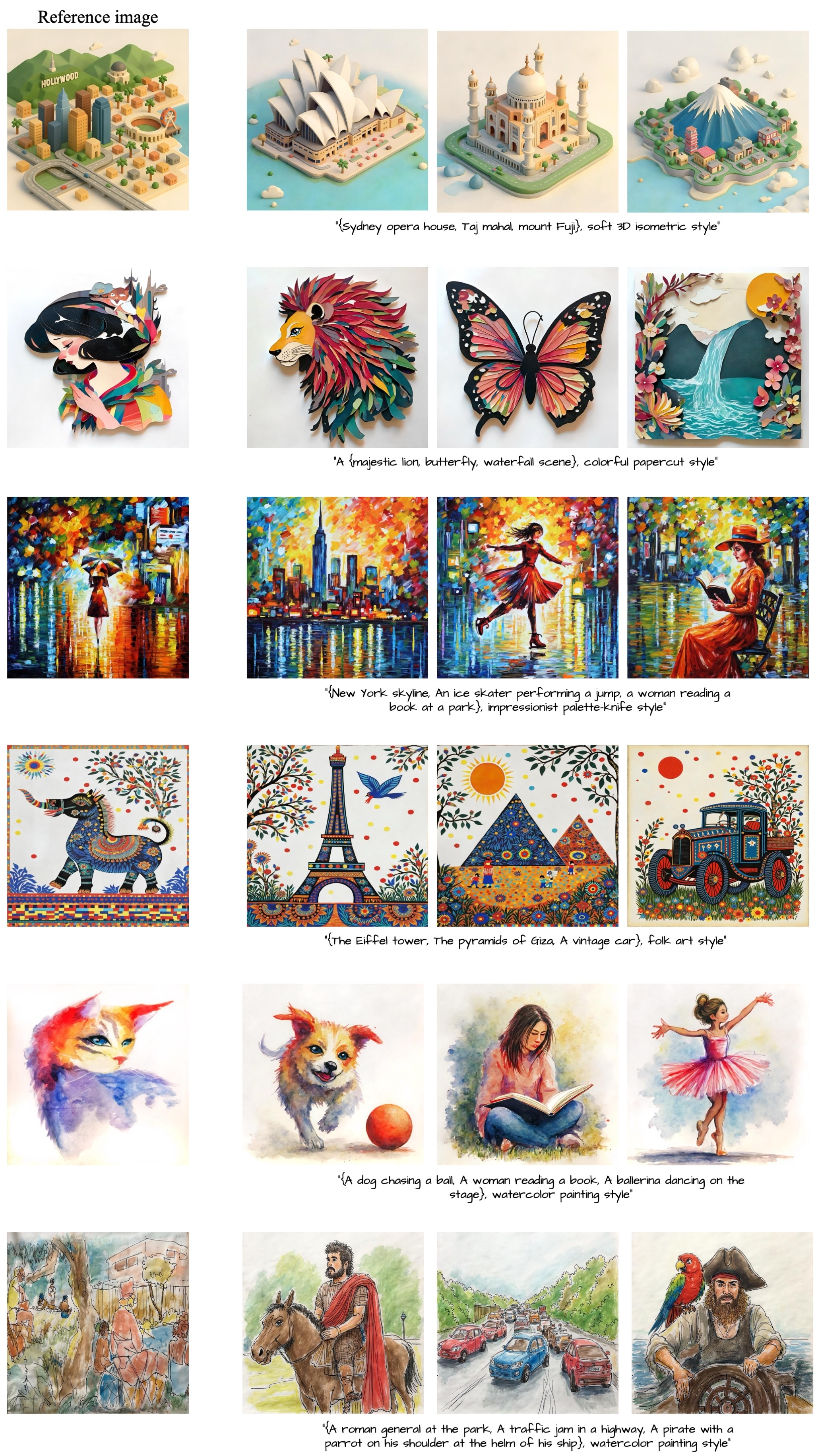}
    \end{overpic}
    \caption{\textbf{Extra style transfer results}
}
\label{fig:extra_2}
\end{figure*}

\begin{figure*}[t]
    \centering
    \begin{overpic}[width=0.7\linewidth]
    {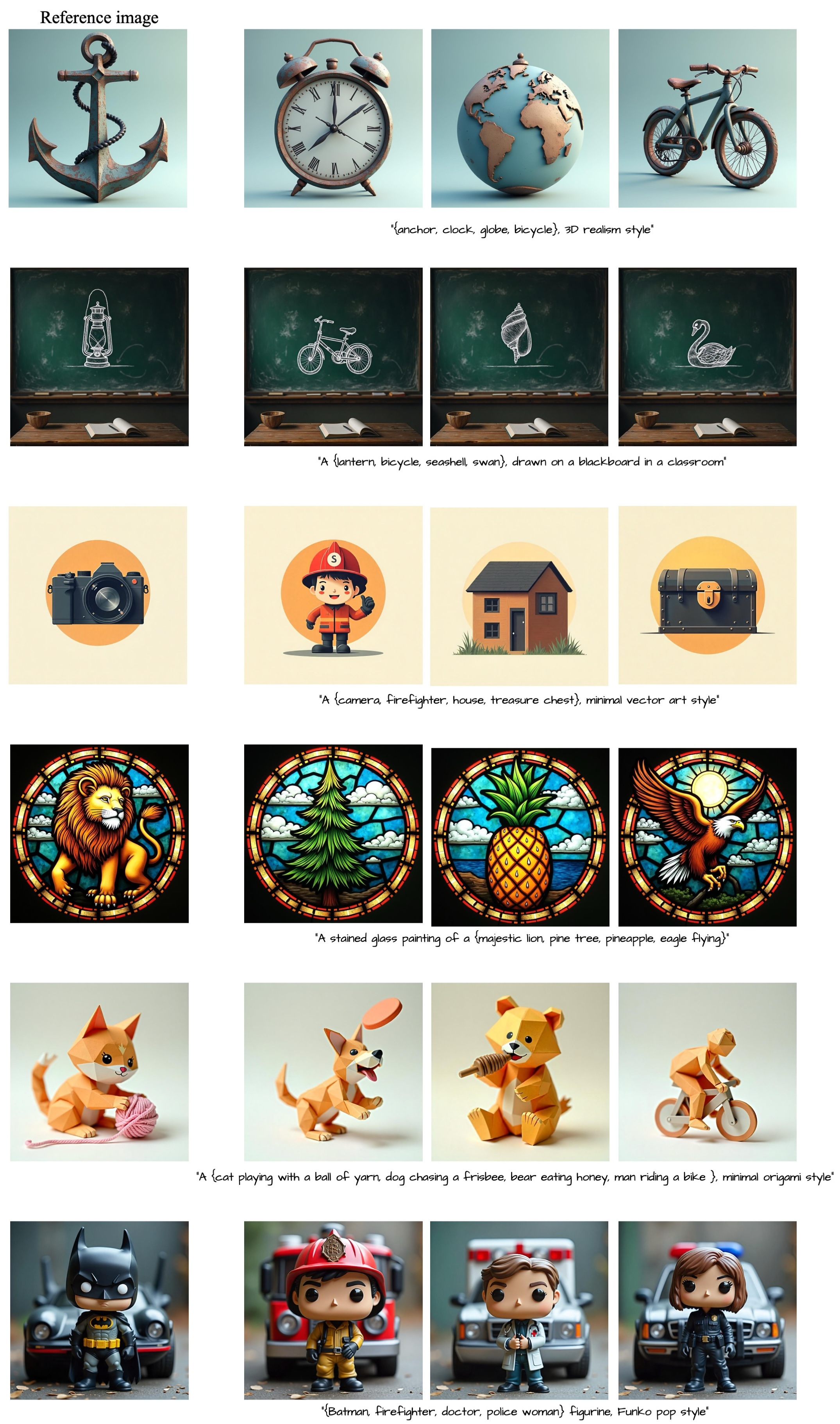}
    \end{overpic}
    \caption{\textbf{Extra style-aligned generation results}
}
\label{fig:extra_3}
\end{figure*}

In \Cref{fig:extra_1} and \Cref{fig:extra_2}, we present additional style transfer results. Our method faithfully transfers the reference style while preserving alignment with the target text prompts. \Cref{fig:extra_3} shows results for style-aligned generation, where the goal is to generate a set of images that are stylistically consistent; that is, the reference image is generated alongside the target images.

\begin{figure*}[t]
    \centering
    \begin{overpic}[width=0.8\linewidth]
    {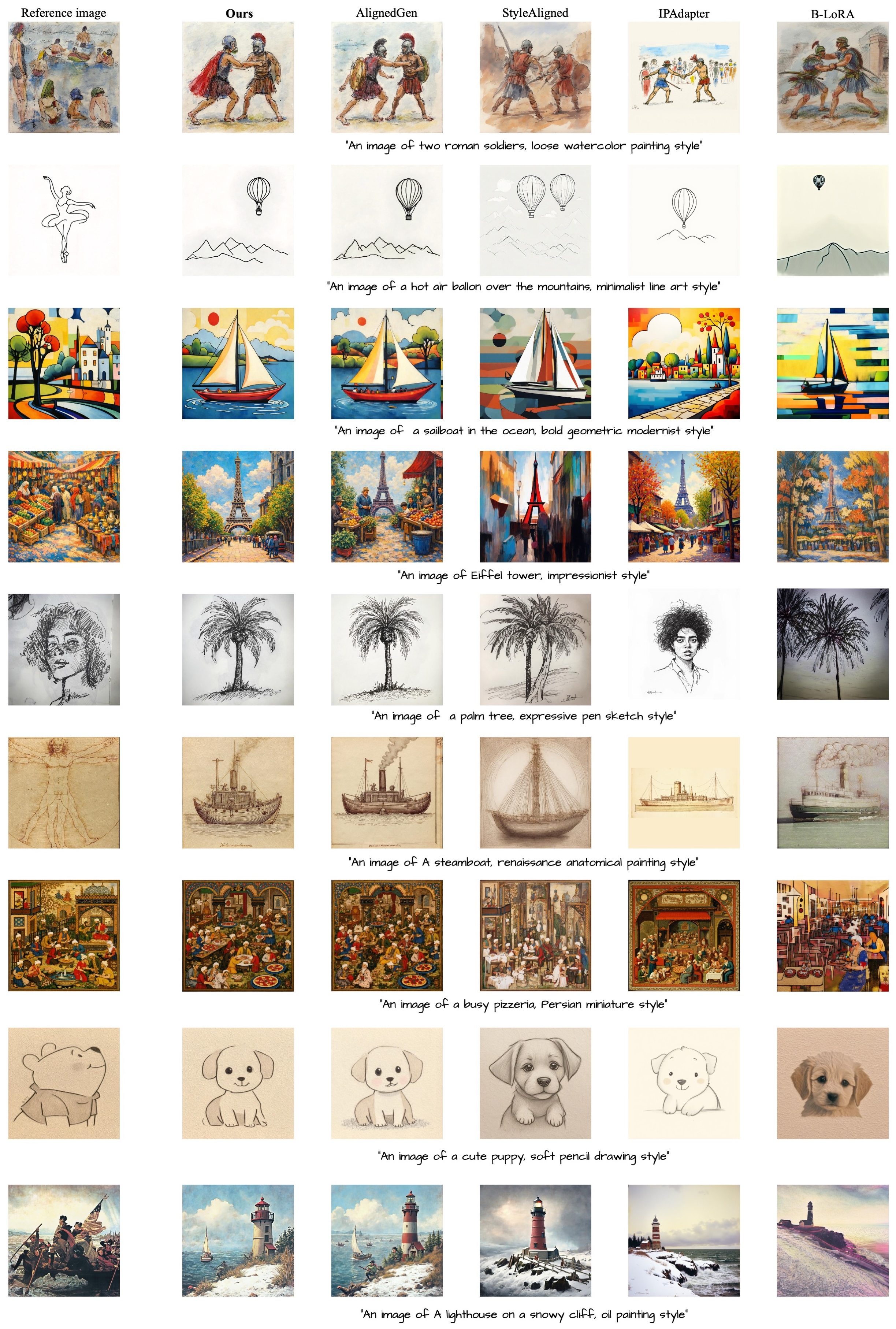}
    \end{overpic}
    \caption{\textbf{Magnified results of \Cref{fig:visual_comparison}}
}
\label{fig:mag_1}
\end{figure*}

\begin{figure*}[t]
    \centering
    \begin{overpic}[width=0.8\linewidth]
    {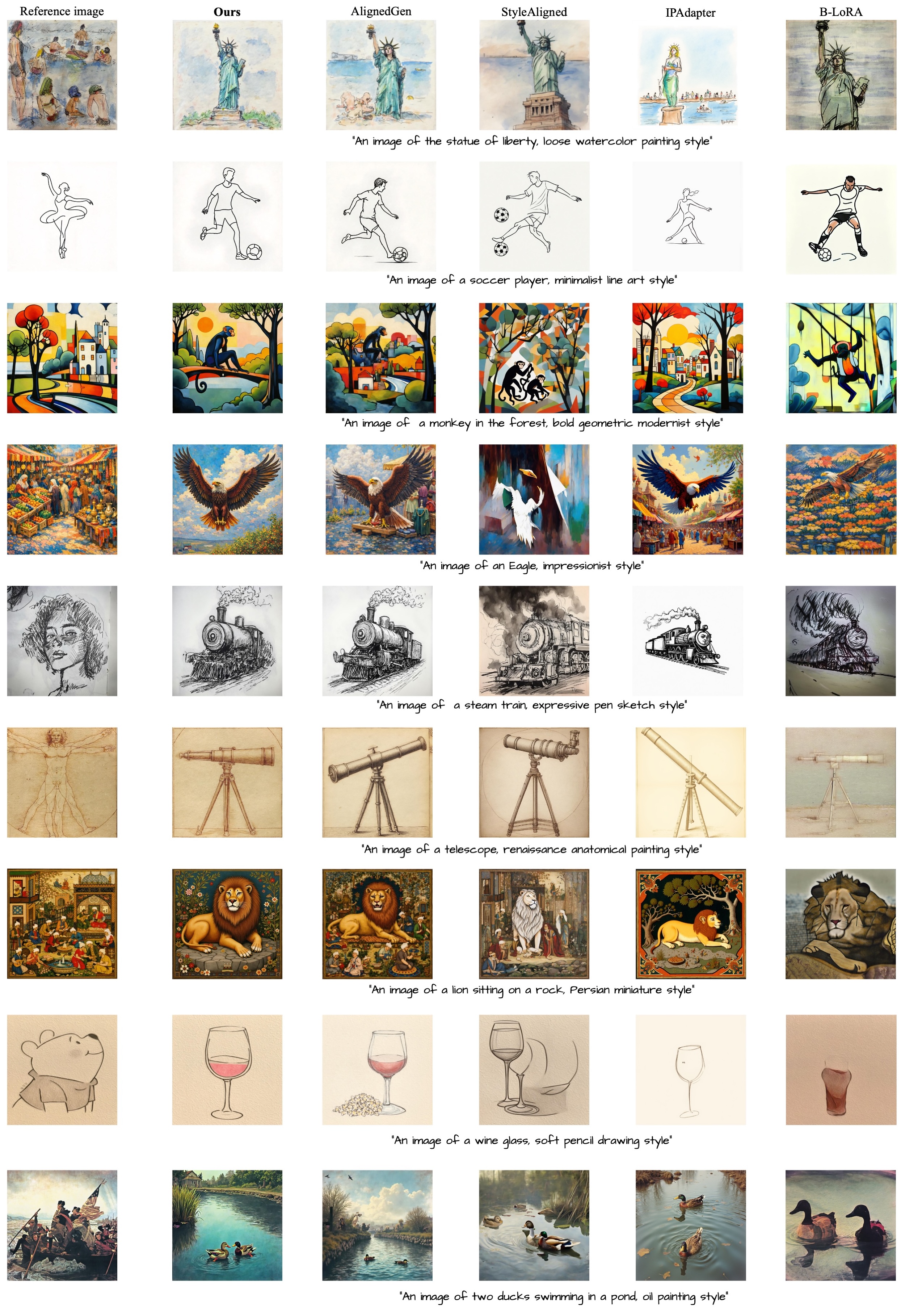}
    \end{overpic}
     \caption{\textbf{Magnified results of \Cref{fig:visual_comparison}}}
\label{fig:mag_2}
\end{figure*}

Finally, we provide magnified views of \Cref{fig:visual_comparison} in \Cref{fig:mag_1} and \Cref{fig:mag_2} for improved visibility. Notably, our method is the only approach that consistently transfers style without introducing unwanted content leakage from the reference.

\end{document}